\documentclass[twocolumn]{aastex631}

\begin{document}

\title{Tracing the Propagation of Shocks in the Equatorial Ring of SN 1987A Over Decades with the Hubble Space Telescope}

\correspondingauthor{Christos Tegkelidis}
\email{cteg@kth.se}

\author[0009-0007-3359-5767]{Christos Tegkelidis}
\affiliation{Department of Physics, KTH Royal Institute of Technology, The Oskar Klein Centre, AlbaNova, SE-106 91 Stockholm, Sweden}

\author[0000-0003-0065-2933]{Josefin Larsson}
\affiliation{Department of Physics, KTH Royal Institute of Technology, The Oskar Klein Centre, AlbaNova, SE-106 91 Stockholm, Sweden}

\author[0000-0001-8532-3594]{Claes Fransson}
\affiliation{Department of Astronomy, Stockholm University, The Oskar Klein Centre, AlbaNova, SE-106 91 Stockholm, Sweden}




\begin{abstract}

The nearby SN 1987A offers a unique opportunity to investigate the complex shock interaction between the ejecta and circumstellar medium. We track the evolution of the optical hotspots within the Equatorial Ring (ER) by analyzing 33 Hubble Space Telescope imaging observations between 1994 and 2022. By fitting the ER with an elliptical model, we determine its inclination to be $ 42.85 \pm 0.50^{\circ}$ with its major axis oriented $ -6.24 \pm 0.31^{\circ}$ from the west. We identify 26 distinct hotspots across the ER, with additional ones emerging over time, particularly on the west side. The hotspots initially show high velocities ranging from $390$ to $1660 \ \rm km \ s^{-1}$, followed by a deceleration phase around day $\sim 8000$. Subsequent velocities vary from $40$ to $660 \ \rm km \ s^{-1}$. The light curves of the hotspots reach maxima between 7000 and 9000 days, suggesting a connection with the deceleration. Many spots are spatially resolved and show elongation perpendicular to the direction of motion, indicative of a short cooling time. To explain these results, we propose that each hotspot comprises dense substructures embedded in less dense gas. The initial velocities are then phase velocities, the break occurs when the blast wave leaves the ER, while the late velocities reflect the propagation of radiative shocks in the dense substructures. We estimate that the dense substructures have a volumetric filling factor of $\sim0.3 \left(  n_{\mathrm{e}}/10^{6}\ \mathrm{cm^{-3}} \right)^{-2} \%$ and a total mass of $\sim0.24 \left(n_{\mathrm{e}}/10^{6}\ \mathrm{cm^{-3}} \right)^{-1}\times10^{-2}\ \mathrm{M_{\odot}}$.

\end{abstract}

\keywords{Supernova remnants (1667) --- shocks (2086) -- core-collapse supernovae (304)}


\section{Introduction} \label{sec:Introduction}

Supernova (SN) 1987A is one of the most extensively studied SNe in the history of mankind. This type II SN, situated in the Large Magellanic Cloud (LMC), approximately 50 kiloparsecs away, was initially observed on February 23, 1987 \citep{McCray_1993, McCray_2016}. Its relatively short distance, compared to other SNe, presented a unique opportunity for in-depth observations and study of its evolutionary stages. The SN remnant (SNR) it left behind, resulting from the interaction between the blast wave driven by expelled material and the circumstellar medium (CSM), can offer valuable clues about the star's life, composition, and the shock physics involved.

The environment surrounding SN 1987A is characterized by a circumstellar nebulosity. A notable feature within this nebula is the triple ring structure, consisting of two outer rings (OR) and one inner, equatorial ring (ER), all exhibiting axisymmetry, nearly circular shapes, and inclinations in the range $38^{\circ} -45^{\circ}$ \citep{Tziamtzis_2011, Crotts_2000}. Studies based on the kinematics of the unshocked rings suggest that these structures formed approximately 20,000 years before the SN explosion \citep{Crotts_2000}. Several models have been proposed to explain the creation of the rings, including a binary merger \citep{Morris_2007, Morris_2009}.
meaning
A binary merger also stands out as the leading candidate for explaining the properties of the progenitor of SN 1987A, Sanduleak -69 202. This was a blue supergiant (BSG) with an estimated mass of approximately $\sim 15-20$ $\mathrm{M}_{\odot}$ \citep{West_1987, White_1987, Kirshner_1987, Fransson-Kozma, Smartt_2009, Utrobin_2015, Utrobin_2019}. Many studies have found that binary merger models are favored to explain the properties of the progenitor star itself, matter mixing in the SN, as well as the properties of the remnant \citep{Menon_2017, Menon_2019, Ono_2020, Orlando_2020, Utrobin_2021}.

The initial flash of soft X-ray and UV photons from the SN shock breakout ionized the CSM, leading to the discovery of the rings several months after the explosion \citep{Crotts_1989}. Since then, the photoionized gas in the rings has been gradually fading due to recombination \citep{Fransson_1989, Lundqvist_1996, Mattila_2010}. Recombination is an ongoing process for the outer rings (OR), which have not yet experienced shock interactions. However, a different scenario unfolded for the ER when the ejecta-driven blast wave initiated interaction.

The innermost region of the ER contained gas clumps with higher densities than the surrounding medium \citep{sn1987_review}. Approximately 5000 days post-explosion, the blast wave collided with this dense matter. This collision gave rise to a slower forward shock and the creation of a reflected shock. The forward shock advanced into the clumps, while the reflected shock moved toward the expanding debris, leading to the formation of a reverse shock \citep{sn1987_review}. The first indication of this collision emerged in 1997 with the appearance of the first optical hotspot \citep{Sonneborn_1998}. Subsequently, multiple such hotspots became visible across the ER, as these dense structures underwent shock interactions and were accelerated by the blast wave \citep{Lawrence_2000}. Their velocities are indicative of the typical speed of transmitted shocks driven into the dense clumps \citep{Larsson_2019}.

A wide range of shock velocities is transmitted through the dense material, determined by the blast wave velocity, clump density, and geometry \citep{sn1987_review}. However, the largest contribution of optical emission is produced from shocks that had time to become radiative \citep{pun_2002}. Optical emission line widths from the ER suggest velocities between $200$ and $500\ \mathrm{km\ s^{-1}}$ \citep{Groeningsson_2008_02}. These line profiles predominately stem from the recombining postshocked gas in the ER, reflecting the projected velocities along the line of sight of radiative shocks transmitted through the dense obstacles \citep{Groeningsson_2008}. Moreover, early narrow emission lines following the initial flash ionization implied a density range for the unshocked gas in the ER of order $10^{3}-10^{4}\ \mathrm{cm}^{-3}$ \citep{Lundqvist_1996}. In contrast, nebular diagnostics based on optical and UV narrow emission lines imply densities of the shocked gas in the ER in the range $10^{6}-10^{7}\ \mathrm{cm}^{-3}$ \citep{Groeningsson_2008, pun_2002}.

Following the interaction of the blast wave with the ER, the optical flux started to increase until $\sim 8000$ days \citep{Fransson_2015}. Since then, the optical emission began to decrease, indicating that the blast wave has moved past the dense clumps and that the clumps are getting destroyed \citep{Fransson_2015}. A similar evolution is observed for both the near- and mid-infrared light curves, which peaked at $8500$ days \citep{Arendt_2016, Kangas_2023}, as well as the soft X-rays, which leveled off at $\sim 9500$ days \citep{Frank_2016} and started to decay after $\sim$ 11,000 days \citep{Ravi_2024}. In contrast, the hard X-rays and radio light curves continued to increase and their similarities indicate that they trace lower-density regions of the ER \citep{Zanardo_2010, Arendt_2016, Ravi_2024}.

The proper motion of the hotspots measured in images can provide new insights about the nature of the shocks and CSM, complementing the information obtained from spectra and light curves. In this paper, we use HST imaging in the R-band over the last 28 years to track the temporal and spatial evolution of these hotspots. Our analysis involves measuring hotspot velocities, detecting potential changes in velocities over time or across the ER, monitoring flux evolution since emergence, and determining if certain hotspots are spatially resolved for a more accurate assessment of their physical size. 

This paper is organized as follows. First, we present the observations in Section \ref{sec:Observations and Data Reduction}, while the analysis and the results are described in Section \ref{sec:Analysis and Results}. Section \ref{sec: Discussion} provides a discussion and finally the conclusions are summarized in Section \ref{sec: Summary}.

\section{Observations and Data Reduction} \label{sec:Observations and Data Reduction}

\begin{figure*}[htb!]
    \plotone{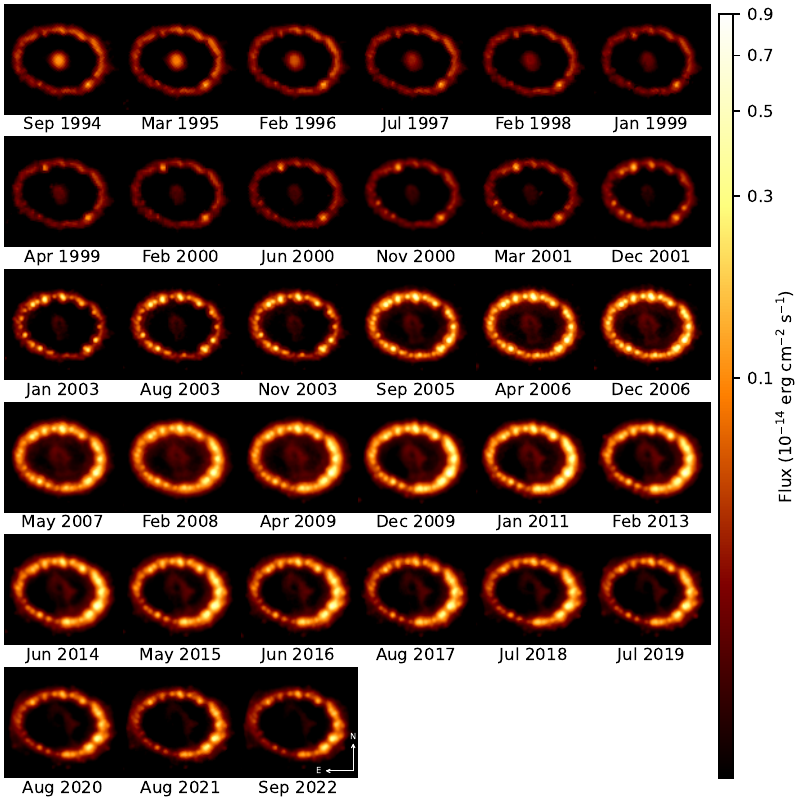}
    \caption{All HST R-band images used in the study, ranging from 1994 to 2022. The color scale is logarithmic, with the maximum set to the brightest hotspot in December 2009 to showcase the evolution of the ER. The FOV of each image is $2\farcs30  \times 2\farcs15$. The spot in the southwest at approximately 5 o'clock is a star. We refer the reader to \cite{Larsson_2019} and \cite{Sophie_2024} for versions of these images that have color scales set to highlight the ejecta inside the ER. \label{fig:ring evolution}}
\end{figure*}

Table \ref{tab:observations} summarizes the details of all utilized observations. The analysis is based on 33 HST imaging observations \footnote{Two additional observations are available in the archive (from 2005 September 28 and 2006 April 29), but were excluded due to strong artifacts in the images.} in various broad R-band filters since 1994 shown in Figure \ref{fig:ring evolution}. The observations between 1994 and 2001, as well as between 2007 and 2009, were conducted by the WFPC2 instrument using the UV/Visible channel (UVIS) camera with the F675W filter. The measurements between 2007 and 2009 were affected by the degradation of the Charge Transfer Efficiency (CTE) of the instrument, leading to higher uncertainties in the measured fluxes \citep{Larsson_2019}. Observations between 2003 and 2006 were performed by the ACS instrument using the High Resolution Channel (HRC) with the F625W filter. Observations from 2009 and onward were performed by the WFC3 instrument using the UVIS channel with the F625W filter. The emission in all images is strongly dominated by the H$\alpha$ line. The extracted spectrum of the ER is available in Figure 4 in \cite{Groeningsson_2008_02}. The filter width is slightly different for the three detectors. This does not affect the measurements of the hotspot positions, but we correct for it when deriving fluxes. More information about the filters can be found in \cite{Acs, WFPC2, WFC3}.

The details of the image processing are discussed in previous papers \citep{Larsson_2011, Fransson_2015, Larsson_2019, Sophie_2024}. We briefly mention that all images were drizzled to combine dithered exposures and remove cosmic rays. All images were drizzled to a scale of $25 \ \rm{mas}$ per pixel and pixel-aligned with each other.

\begin{deluxetable*}{c c c c c c c c c c}[htb]
\tabletypesize{\scriptsize}
\tablewidth{0pt} 
\tablecaption{HST Observations of SN 1987A in the R-band \label{tab:observations}}
\tablehead{
\colhead{Date} & \colhead{Epoch\tablenotemark{a}}& \colhead{Instrument} & \colhead{Filter} & \colhead{Exposure} & \colhead{Date} & \colhead{Epoch\tablenotemark{a}}& \colhead{Instrument} & \colhead{Filter} & \colhead{Exposure} \\
\colhead{(YYYYmmdd)}& \colhead{(days)} & \colhead{} & \colhead{} & \colhead{($\rm s$)} & \colhead{(YYYYmmdd)}& \colhead{(days)} & \colhead{} & \colhead{} & \colhead{($\rm s$)}
}   
\startdata 
1994-09-24 & 2770 & WFPC2 & F675W & 600  & 2007-05-12 & 7384  & WFPC2 & F675W & 2700 \\
1995-03-05 & 2932 & WFPC2 & F675W & 600  & 2008-02-19 & 7666  & WFPC2 & F675W & 1600\\
1996-02-06 & 3270 & WFPC2 & F675W & 600  & 2009-04-29 & 8101  & WFPC2 & F675W & 1600\\
1997-07-10 & 3790 & WFPC2 & F675W & 600  & 2009-12-12 & 8328  & WFC3  & F625W & 3000\\
1998-02-06 & 4001 & WFPC2 & F675W & 400  & 2011-01-05 & 8717  & WFC3  & F625W & 1140\\
1999-01-07 & 4336 & WFPC2 & F675W & 1220 & 2013-02-06 & 9480  & WFC3  & F625W & 1200\\
1999-04-21 & 4440 & WFPC2 & F675W & 400  & 2014-06-15 & 9974  & WFC3  & F625W & 1200\\
2000-02-02 & 4727 & WFPC2 & F675W & 400  & 2015-05-24 & 10317 & WFC3  & F625W & 1200\\
2000-06-16 & 4862 & WFPC2 & F675W & 400  & 2016-06-08 & 10698 & WFC3  & F625W & 600\\
2000-11-13 & 5013 & WFPC2 & F675W & 2400 & 2017-08-03 & 11119 & WFC3  & F625W & 1200\\
2001-03-23 & 5142 & WFPC2 & F675W & 500  & 2018-07-08 & 11458 & WFC3  & F625W & 1200\\
2001-12-07 & 5401 & WFPC2 & F675W & 800  & 2019-07-22 & 11837 & WFC3  & F625W & 1200\\
2003-01-05 & 5796 & ACS   & F625W & 800  & 2020-08-06 & 12218 & WFC3  & F625W & 1160\\
2003-08-12 & 6014 & ACS   & F625W & 480  & 2021-08-21 & 12598 & WFC3  & F625W & 1080\\
2003-11-28 & 6122 & ACS   & F625W & 800  & 2022-09-05 & 12978 & WFC3  & F625W & 1080\\
2005-09-26 & 6790 & ACS   & F625W & 12000 \\
2006-04-15 & 6991& ACS    & F625W & 1200 \\
2006-12-06 & 7226 & ACS   & F625W & 1200 \\
\enddata
\tablenotetext{a}{Since 1987-02-23.}
\end{deluxetable*}

\section{Analysis and Results} \label{sec:Analysis and Results}

We adopt a distance to SN 1987A of 49.6 kpc \citep{2019Natur.567..200P}. At this distance, $1''$ corresponds to $\sim 0.24 \ \rm pc$ ($\sim 7.42 \times 10 ^{17} \ \rm cm$). In the following analysis, the hotspots (also referred as spots) are described as point-like sources and are assumed to lie in the plane of the ER. We use the term ``radial distances'' to refer to the distances between the hotspots and the center of the ER. The latter was taken as the favored position from \cite{Alp_2018}, which was estimated using a method similar to the one described in Section \ref{subsec: Inclination Correction}, involving imaging observations in both R- and B-bands. We present the methodology used to identify and fit the hotspots in Section \ref{subsec:Identification and Fitting of Hotspots} and the correction for the inclination in Section \ref{subsec: Inclination Correction}. We describe the radial and angular evolution of the spots in Section \ref{subsec: Emergence and Movement of the Spots}, followed by the flux evolution, the velocities, and the sizes of the spots in Sections \ref{sub: Flux Evolution}-\ref{subsection: Sizes}, respectively. Appendix \ref{Appendix: Fitting} provides the details of the different fitting functions used, while the method used for the inclination correction is included in Appendix \ref{Appendix: Inclination}. Finally, the radial fitting of the individual hotspots can be found in Appendix \ref{Appendix: Radial Fitting of the Spots}.

\subsection{Identification and Fitting of Hotspots}\label{subsec:Identification and Fitting of Hotspots}

\begin{figure*}
    \plotone{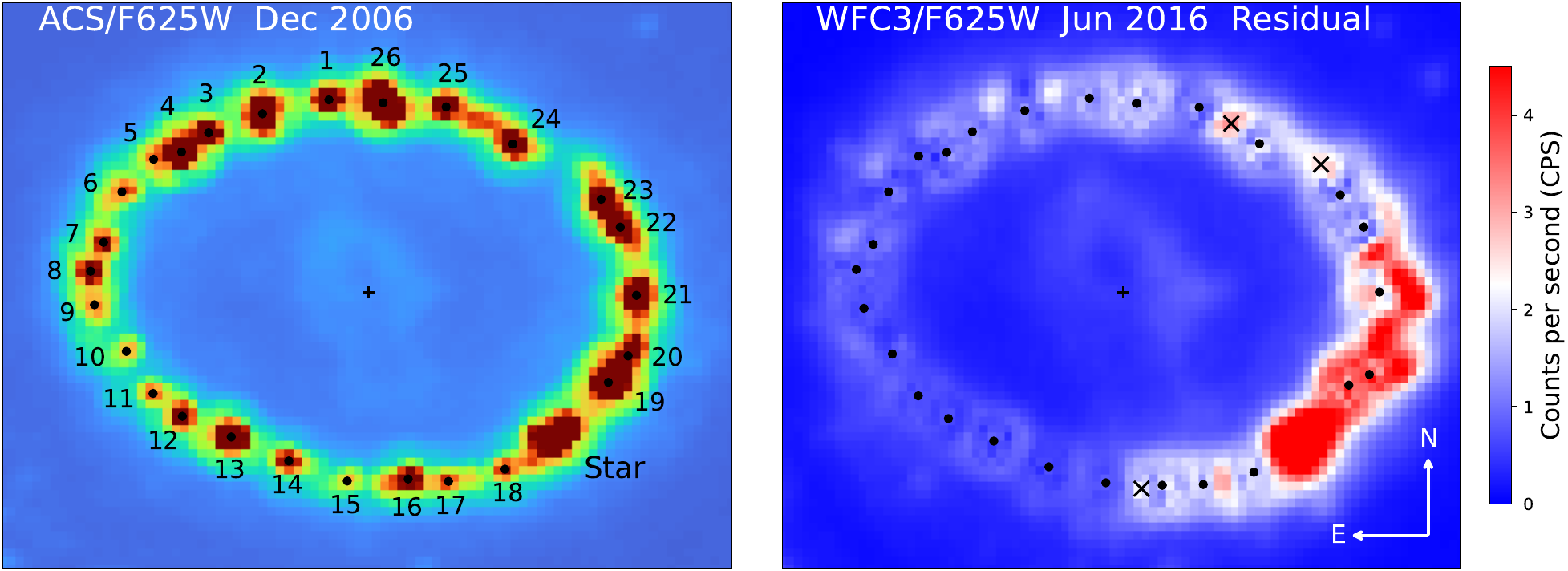}
    \caption{Left panel: ACS/F625W image from 7226 days after the explosion (2006 December 6) showing all the identified hotspots in the ER as dots. The numbering of the spots is done anti-clockwise from the north. Spot 5 is the one spot detected from the residual image at the second iteration. The center of the ER is marked by a black cross. The FOV is $2\farcs10 \times1\farcs73$. Right panel: WFC3/F625W residual image from 10,698 days (2016 June 8) after the subtraction of all 26 fitted spots from the original image.
    The black dots represent the centroids of the 26 hotspots while the black crosses correspond to the 3 additional spots detected at the second iteration that were not included in our analysis. Red means brighter. Note the additional emission in the west which becomes increasingly brighter in all residual images after 2015. The FOV is $2\farcs30 \times2\farcs13$. \label{fig:spots}}
\end{figure*}

To determine the positions of the hotspots, we employed the versatile PSF photometry and detection algorithms provided by \texttt{photutils} \citep{photutils_1.9}. We selected a 2D elliptical Gaussian function to capture potential elongation in the hotspots. The free parameters of the model are the centroid $x_0$, $y_0$, the x and y widths $\sigma_x$, $\sigma_y$, and the counterclockwise angle of rotation $\theta$ of the blob. The details of the model can be found in Appendix \ref{Appendix: Fitting}. We opted for the Levenberg–Marquardt optimization algorithm. 

The sources were fitted over a 125$\times$125 mas\(^{2}\) region, using an aperture radius of 62.5 mas for initial flux estimation. To address challenges posed by overlapping sources, we initiated the grouper parameter, allowing for the simultaneous fitting of sources close to each other. We varied the maximum distance for which sources were fitted together between 150 and 220 mas, ultimately selecting 170 mas for all fits for consistency. In addition, we enabled the local background estimation during the fitting process, which calculates the local statistics in a circular annulus aperture around each source with sigma clipping. The inner radius of the background aperture was allowed to range between 87.5 to 125 mas, with the outer radius extending by 62.5 to 87.5 mas beyond. The final selected background region for each spot was based on the the absolute value of the sum of the fit residuals. Although there were instances of grouped sources with overlapping regions where different local backgrounds were applied, this is unlikely to affect the fits, as the background variations between adjacent spots were small, typically less than 12\%. The impact of different background and grouping configurations on both the fitted centroids and source widths was comparable to the size of their respective statistical uncertainties, with variations of less than 3\% for the former and approximately 5\% for the latter. The largest systematic differences were observed in the calculated fluxes, with a median value of 10\% and a 90th percentile of 22\%, surpassing reported statistical uncertainties. Unless otherwise stated, the uncertainties associated with the best-fit parameters were calculated as the square root of the diagonal elements of the covariance matrix.

A crucial step in studying the evolution of these hotspots is their identification. The existence of the spots was confirmed using the \texttt{photutils} \texttt{DAOStarFinder} tool, which searches for Gaussian-like sources above a specified threshold and a size resembling the defined Gaussian kernel. We specifically searched for slightly elliptical sources with an axis ratio of 0.9, although the same sources were identified when a symmetric kernel was employed. To ensure consistency in tracking these sources over the years, we initiated the detection algorithm on the ACS/F625W image from 7226 days after the explosion (2006 December 6), which offers the highest spatial resolution and corresponds to a period when the majority of the spots exhibited increased brightness \citep{Fransson_2015}. The regions containing the star at approximately 5 o'clock (see Figure \ref{fig:spots}) and the inner ejecta were masked. The detection threshold was defined as the median value of the image section encompassing the ER and an additional three standard deviations to avoid detecting background peaks. To further validate the accuracy of the process, we experimented with varying the size of the kernel, ranging from 2.5 to 4.5 pixels, and its ellipticity. Across all cases, the same number of spots were identified.

This procedure resulted in the detection of 25 hotspots. The residual images, generated by subtracting the 25 fitted spots, revealed potential faint source-like peaks. To identify additional faint spots, we applied the detection algorithm on the ACS/F625W residual image from day 7226. This second iteration led to the discovery of four additional hotspots, of which only one was considered in our analysis (spot 5 in Figure \ref{fig:spots}). The remaining three were excluded due to a lack of systematic behavior. The final 26 hotspots are shown in Figure \ref{fig:spots}.

The detected hotspot coordinates served as the initial positional parameters for their fitting in all observations from day 5796 onward. However, a different approach was necessary for images before day 5796, as accurately fitting the majority of spots in these early observations was challenging. This difficulty stemmed from the fact that most of the spots detected in the 2006 December 6 image had not yet emerged in these earlier observations. To address this, we ran the detection algorithm on images taken between day 2770 and day 5401 (WFPC2), subsequently fitting only the sources that were detected in each instance and had coordinates close to the 25 spots recorded in the ACS/F625W image from 2006 December 6. This method simplified the task of tracking when a spot became visible in the data set since the emergence of a spot is not a straightforward occurrence and varies throughout the ER. This is discussed further in Section \ref{subsec: Emergence and Movement of the Spots}. The systematic uncertainties in hotspot coordinates, resulting from different background and maximum grouping distance configurations, were slightly larger for WFPC2 images compared to those obtained after day 5769, but remained under 3\%. As illustrated in the right panel of Figure \ref{fig:spots}, the residual images supported the appearance of new spots in the western part of the ER around day 10,000 and beyond. However, we did not include these in our analysis due to the limited data points available for describing their radial evolution. 

\subsection{Inclination Correction}\label{subsec: Inclination Correction}

The ER is expected to be nearly circular but appears elliptical in images due to its inclination to the plane of the sky \citep{Tziamtzis_2011}. Consequently, the distribution of the hotspots becomes contracted, affecting the measured distances and, thereby, the calculated velocities. We account for this inclination effect by applying a matrix transformation defined by the inclination and orientation angles of the ER. The orientation here is the angle of the major axis measured counterclockwise on the image plane from the west. Under this transformation, hotspots on the major axis should remain unchanged, while the rest should increase their radial distance. The orientation and inclination angles were estimated by two different methods.
 
The first method involved fitting ellipses to the main 26 spot positions in each image between day 6122 and day 12,598. Early observations were excluded due to a limited number of clearly visible spots, which rendered the fits highly inaccurate. The most recent observation was also omitted for a similar reason, as some of the main spots could not be accurately fitted.

The second method was considered to address the systematic uncertainty introduced by treating the ring as a collection of point sources rather than a diffuse structure. For that reason, a diffuse elliptical annulus with a Gaussian radial profile was fitted to all observations between day 5796 and day 12,978 (Equation \ref{eq:gaus ring}). This annulus is allowed to rotate on the image plane and has a sinusoidal surface brightness that varies as a function of the azimuth. This method describes the ER as a unified entity rather than a set of individual point sources and provides an independent measurement of its inclination and orientation.

Table \ref{tab:geometry} provides a summary of the arithmetic means for the relevant best-fit parameters obtained from both methods. The reported uncertainties correspond to the standard deviations of the sample of the best-fit ellipses and diffuse annulus, respectively. The results obtained from the two distinct methods agree within the 3 sigma uncertainties, except for the orientation angle. The inclination angle can be compared with the results obtained by \cite{Tziamtzis_2011}, where an inclination of around $43^{\circ}$ was reported. Due to the non-constant radial distance of the hotspots, we choose to adopt the results from the diffuse elliptic annulus method for the analysis. However, these slight differences are not expected to significantly impact the overall findings.

\begin{deluxetable}{ccc}[htb]
\tabletypesize{\scriptsize}
\tablewidth{0pt} 
\tablecaption{Geometry of the ER \label{tab:geometry}}
\tablehead{
\colhead{Parameters} & \colhead{Ellipse \tablenotemark{a}} & \colhead{Diffuse Annulus \tablenotemark{b}}} 
\startdata
{Inclination (deg)}    & 43.51 $\pm$ 0.38  & \textbf{42.85} $\mathbf{\pm}$ \textbf{0.50}\\
{Orientation (deg)}    & -7.84 $\pm$ 0.19  & \textbf{-6.24} $\mathbf{\pm}$ \textbf{0.31} \\
{Avg. Axis ratio}      & 0.725 $\pm$ 0.005 & 0.732 $\pm$ 0.006 \\
\hline
{Major semiaxis (mas)} & 813.96 $\pm$ 8.55 & 821.98 $\pm$ 20.48 \\
{Minor semiaxis (mas)} & 589.42 $\pm$ 6.15 & 592.48 $\pm$ 23.75 \\
{Axis ratio}           & 0.724 $\pm$ 0.001 & 0.721 $\pm$ 0.023\\
\enddata
\tablecomments{The values used in the analysis are indicated in bold. The angles are measured counterclockwise from the west. The first 3 rows, the inclination, orientation, and axis ratio, are averages of all the epochs fitted: 6122-12,598 days for the Ellipse method and 5796-12,978 days for the Diffuse Annulus method. As the ER is expanding, we report the the semi-major and -minor axes as measured from the latest observation only (WFC3/F625W on day 12,978) with uncertainties taken from the covariance matrix. The resulting axis ratio is also provided in the final row. Note that the fit from the last observation (day 12,978) was not included in the averages in the first 3 rows for the Ellipse method.}
\tablenotetext{a}{Ellipse fitted to the 26 hotspot centroids.}
\tablenotetext{b}{Elliptical annulus with a Gaussian radial profile (Apendix \ref{Appendix: Fitting}).}
\end{deluxetable}

\subsection{Emergence and Movement of the Spots}\label{subsec: Emergence and Movement of the Spots}

\begin{figure*}[htb]
    \plotone{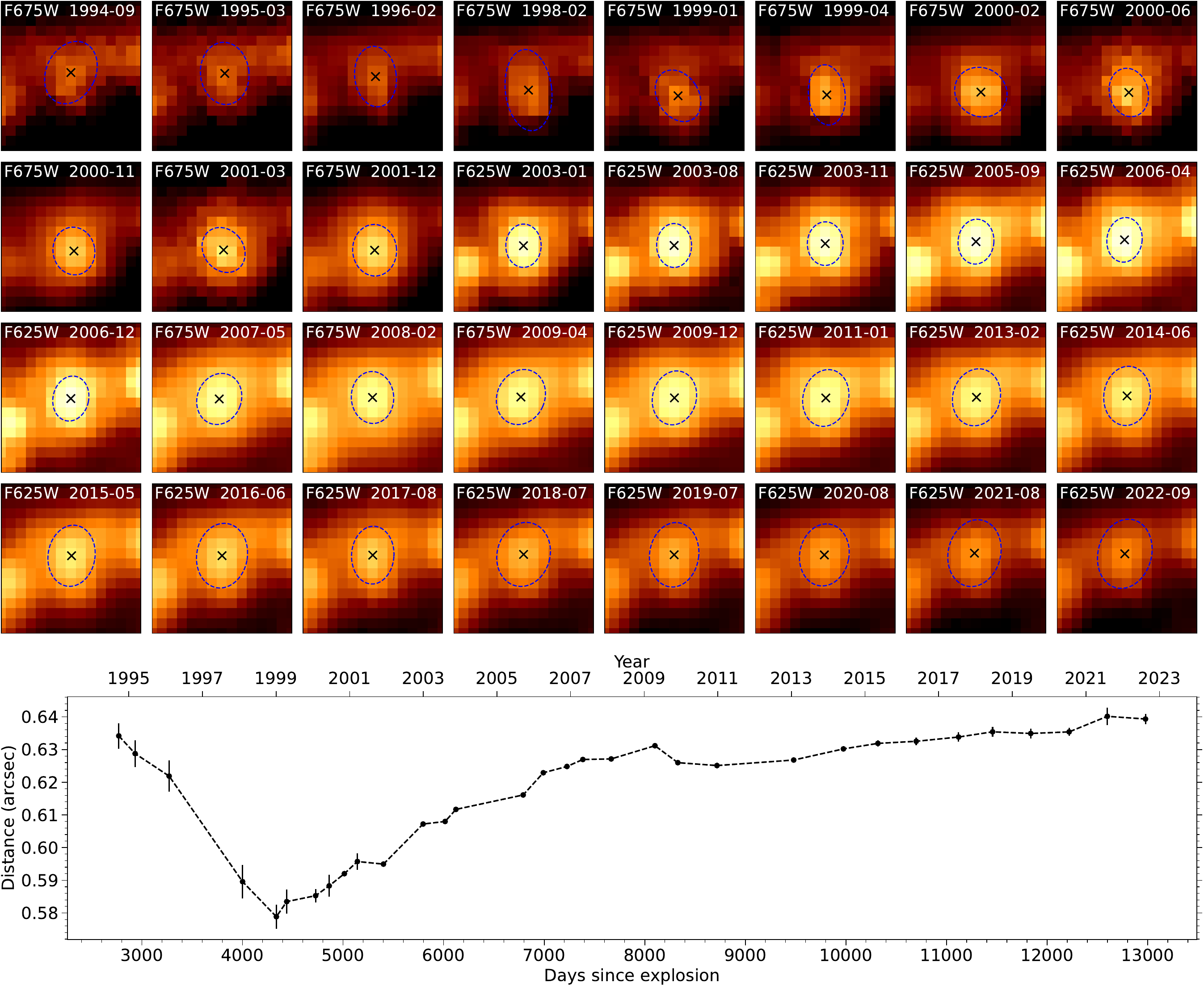}
    \caption{Time evolution of spot 2. The upper panels display the region of spot 2 (as defined in Figure \ref{fig:spots}) for all observations. The cross marker indicates the best-fit centroid of the detected source in that region, while the dashed ellipses showcase the orientation as well as the major and minor FWHM of the spot. The FOV of the panels is $350\times 375 \ \rm mas^{2}$. In the lower panel, the radial distance of the detected source from the center of SN 1987A is depicted over time. Initially, the algorithm identifies and fits a source at the outer edge of the ER. Over time, as the blast wave encounters the ER, emission from the inner part of the ER dominates, causing a drop in the radial distance. Eventually, a minimum is reached around the year 1999, which we identify as the activation time of the spot. Subsequently, the radial distance increases systematically. The same general behavior is observed for the other spots.\label{fig:spot 2 detection}}  
\end{figure*}

\begin{figure*}[htb]
\plotone{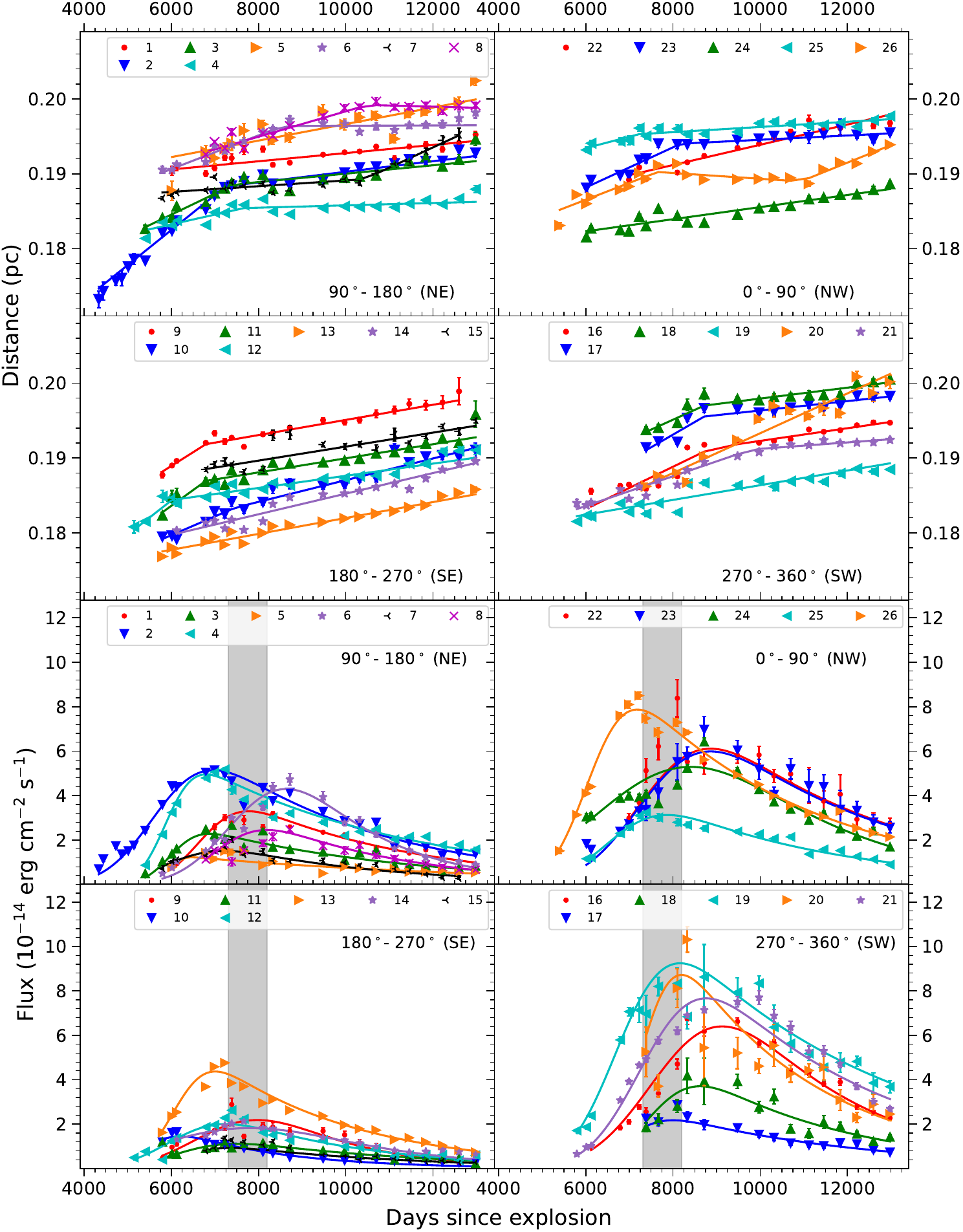}
\caption{Time-evolution of the hotspots. The hotspots are split into 4 different quadrants (northeast (NE), northwest(NW), southeast (SE), and southwest (SW)) based on their position angles. Upper panels: Time evolution of the radial distances of the different spots. The distances are inclination corrected and measured from the center of SN 1987A reported by \cite{Alp_2018}. The solid lines represent the best fits for the data after the emergence of the different spots used for the velocity estimations. Lower panels: Flux evolution of the individual hotspots in the R-band. The grey band represents the epochs of the WFPC2 measurements significantly affected by the CTE loss. To compensate for this, the fluxes of these measurements were multiplied by a factor of 1.25. The solid lines give the exponentially modified Gaussian best-fits to the data after the emergence of the different spots. 
\label{fig:radial evolution}}
\end{figure*}

Figure \ref{fig:spot 2 detection} shows the radial evolution of spot 2 as an example to illustrate the overall trend of how the hotspots emerge. At early observations, the detection algorithm occasionally picks up substructures in the diffused nebulosity of the ER, contributing to the less systematic nature of the evolution of the position in these instances. Following the interaction with the blast wave, the clumps are activated and increase their optical emission. The activation time varies for each spot, manifesting as a radial drop between 4000 and 6000 days. We deem a hotspot to have fully emerged once its radial distance reaches a local minimum and starts to systematically increase thereafter. Figure \ref{fig:radial evolution} illustrates the radial and flux evolution of all hotspots after their activation. Notably, hotspots in the northeast part of the ER (spots 26, 2, 3, 4, 12) are activated first, occurring before 5500 days, while hotspots in the south and west parts of the ER (spots 14-18, 20, 22) are the last to emerge after 6000 days. The remaining hotspots appear at around 6000 days. The emergence time ($t_{\rm start}$) of all the spots is listed in Table \ref{tab:velocities}. 

The evolution of the Position Angles (PAs) of the hotspots, measured anti-clockwise from the west, was also investigated to ensure the validity of the results. Similar to the radial evolution, the PAs exhibit less systematic behavior at early times, followed by a sudden change after the activation time for the majority of the spots. A closer examination revealed that most spots maintained an overall constant angle at post-activation times, with only a few exceptions exhibiting a systematic increase or decrease at a very slow rate (spots 2, 7, 12, 20, 21). This rate of change ranged between $1^\circ$ and $5^\circ$ within a time of $7000$ days. We, therefore, only consider the radial evolution in the following analysis. Table \ref{tab:velocities} provides the PA time averages for all spots observed from day 5796 and beyond, along with the standard deviations of their respective samples.

\subsection{Flux Evolution}\label{sub: Flux Evolution}

The flux densities of the hotspots were obtained from the best-fitting elliptical Gaussians (the ``Flux" parameter in Equation \ref{eq:2dg}). In this process, we also multiplied the counts per second units in the images by the inverse sensitivity of the detectors. To be able to analyze the light curves of the spots, we addressed the variations introduced by using three different instruments (WFPC2, ACS, and WFC3), each with different filters in terms of wavelength coverage and throughput. To correct for these differences, we calculated total fluxes based on the filter profiles and subsequently normalized everything with respect to the WFPC2 filter by using the correction factors derived in \cite{Larsson_2019}. The three WFPC2 measurements taken between 7300 and 8200 days were significantly impacted by CTE degradation of the instrument, resulting in lower measured fluxes \citep{Larsson_2019}. Typically, during these observations, the fluxes were 20\% to 30\% lower than the fluxes recorded just before these instances. This is higher than the $\sim 10\%$ reduction estimated for the ER as a whole \citep{Larsson_2019}, which may be due to differences in the flux- and background levels of individual spots. 

The light curves (Figure \ref{fig:radial evolution}) exhibit an initial increase in emission, followed by a gradual decline resembling an exponential decay. Therefore, we employed the same model as proposed in \cite{Arendt_2020} to characterize the flux evolution of the hotspots. The model is an exponentially modified Gaussian (Equation \ref{eq:flux}), which suggests an initial emission as the shock advances into the ER and a subsequent gradual decay after the shock has passed. The three WFCP2 measurements mostly impacted by the CTE loss were excluded during the fitting process. We opted for weighted least-squares optimization with the Trust Region Reflective algorithm, which is well-suited for constrained problems \citep{2020SciPy-NMeth, TRF_1999}. Since the maximum brightness ($\rm Flux_{\rm max}$) and the time of maximum brightness ($t_{\rm peak}$) are not direct variables of the model and generally lack a closed-form solution, we estimated their uncertainties using Monte Carlo simulations. This involved assuming a multivariate normal distribution specified by the best-fit parameters and covariance matrix obtained from the flux optimization. Figure \ref{fig:radial evolution} and Table \ref{tab:velocities} present the measured fluxes of the spots alongside the best-fit models and relevant parameters that describe them.

\subsection{Velocity Evolution and Correlations}\label{subsec: Velocity Evolution and Correlations}

Assuming that the acceleration due to the shock interactions occurs radially, we measure the expansion velocities by fitting the time evolution of the radial distances. Inspection of the time evolution plots (Figure \ref{fig:radial evolution} and Appendix \ref{Appendix: Radial Fitting of the Spots}) suggests a potential change in velocity for many spots. Considering the number of available observations, we employed segmented regression in the form of straight lines and continuous piecewise functions with either one or two breaks. To avoid identifying spurious changes in slopes and break times ($t_{\rm break}$), we imposed a minimum segment length of 1000 days (approximately 4 data points). The fits were conducted using a weighted least-squares optimization with the Trust Region Reflective algorithm. To improve the likelihood of finding the global optimum, a bootstrap restarting process was implemented, involving bootstrapped resamples of the data to refine initial parameters. A partial F-test at the 0.05 level was used to determine which model better described the time evolution of each spot.

\begin{deluxetable*}{c c c c c c c c c c}[htb]
\tabletypesize{\scriptsize}
\tablecaption{Parameters describing the time evolution of the hotspots\label{tab:velocities}}
\tablehead{
\colhead{Spot} & \colhead{PA} & \colhead{$t_{\rm start}$} & \colhead{$v_{1}$} & \colhead{$v_2$} & \colhead{$t_{\rm break}$} & \colhead{$\rm \chi^{2}_{\nu}$ $/$ ndf\tablenotemark{a} } & \colhead{$t_{\rm peak}$} & \colhead{$\rm Flux_{\rm max}$} & \colhead{$\rm \chi^{2}_{\nu}$ $/$ ndf\tablenotemark{b} } \\
\colhead{ } & \colhead{(deg)} & \colhead{(days)} & \colhead{($\rm{km\ s^{-1}}$)} & \colhead{($\rm{km\ s^{-1}}$)} & \colhead{(days)} & \colhead{} & \colhead{(days)} & \colhead{($10^{-14} \ \rm erg \ cm^{-2} \ s^{-1}$)} & \colhead{ }}
\startdata
{Northeast}& {}              &  {}  & {}              &    {}        &     {}         & {}          & {}             & {}              & {} \\
1 & 100.7 $\pm$ 0.5 & 6014 & 195 $\pm$ 30    &    \nodata   &     \nodata    & 4.53 $/$ 18 & 7721 $\pm$ 94  & 3.29 $\pm$ 0.01 & 3.67 $/$ 13\\
2 & 119.2 $\pm$ 0.4 & 4336 & 1672 $\pm$ 75   & 253 $\pm$ 49 & 7245 $\pm$ 112 & 6.13 $/$ 24 & 6903 $\pm$ 76  & 5.22 $\pm$ 0.01 & 10.73 $/$ 18\\
3 & 133.6 $\pm$ 0.4 & 5401 & 999 $\pm$ 159   & 200 $\pm$ 53 & 7445 $\pm$ 278 & 7.43 $/$ 18 & 6860 $\pm$ 191 & 2.26 $\pm$ 0.03 & 7.29 $/$ 14\\
4 & 141.8 $\pm$ 0.2 & 5401 & 467 $\pm$ 95    & 56 $\pm$ 45  & 7666 $\pm$ 469 & 19.50 $/$ 18& 6808 $\pm$ 112 & 4.97 $\pm$ 0.02 & 13.53 $/$ 14\\
5 & 146.7 $\pm$ 0.3 & 6014 & 394 $\pm$ 63    &    \nodata   &     \nodata    & 5.66 $/$ 17 & 6404 $\pm$ 1229& 1.23 $\pm$ 0.17 & 7.34 $/$ 12\\
6 & 157.2 $\pm$ 0.6 & 5796 & 805 $\pm$ 72    & 8 $\pm$ 72   & 8557 $\pm$ 273 & 2.79 $/$ 17 & 8626 $\pm$ 111 & 4.30 $\pm$ 0.01 & 3.13 $/$ 11\\
7 & 168.9 $\pm$ 0.4 & 5796 & 137 $\pm$ 45    & 965 $\pm$ 295& 10428 $\pm$ 424& 6.55 $/$ 16 & 7240 $\pm$ 229 & 1.53 $\pm$ 0.02 & 5.66 $/$ 13\\
8 & 175.3 $\pm$ 0.2 & 6790 & 606 $\pm$ 69    & -54 $\pm$ 220& 10503 $\pm$ 570& 6.94 $/$ 14 & 8170 $\pm$ 126 & 2.45 $\pm$ 0.01 & 1.11 $/$ 11\\
\hline
{Southeast}& {}              &  {}  & {}              &    {}        &     {}         & {}          & {}             & {}              & {} \\
9  & 183.2 $\pm$ 0.6 & 5796 & 1315 $\pm$ 244 & 361 $\pm$ 44 & 6791 $\pm$ 197 & 3.32 $/$ 16 & 7974 $\pm$ 163 & 2.16 $\pm$ 0.01 & 3.65 $/$ 13\\
10 & 194.6 $\pm$ 0.4 & 5796 & 897 $\pm$ 130  & 518 $\pm$ 58 & 7466 $\pm$ 677 & 2.83 $/$ 17 & 6237 $\pm$ 131 & 1.45 $\pm$ 0.01 & 11.55 $/$ 9\\
11 & 206.4 $\pm$ 0.3 & 5796 & 1505 $\pm$ 352 & 336 $\pm$ 28 & 6791 $\pm$ 227 & 2.82 $/$ 17 & 7563 $\pm$ 157 & 1.11 $\pm$ 0.01 & 2.87 $/$ 13\\
12 & 215.3 $\pm$ 0.4 & 5142 & 1632 $\pm$ 462 & 290 $\pm$ 26 & 6015 $\pm$ 207 & 4.04 $/$ 19 & 7336 $\pm$ 220 & 1.97 $\pm$ 0.02 & 17.61 $/$ 16\\
13 & 228.2 $\pm$ 0.3 & 5796 & 380 $\pm$ 20   &    \nodata   &     \nodata    & 22.75 $/$ 19& 7096 $\pm$ 111 & 4.30 $\pm$ 0.02 & 36.72 $/$ 14\\
14 & 246.0 $\pm$ 0.5 & 6122 & 492 $\pm$ 35   &    \nodata   &     \nodata    & 14.65 $/$ 17& 7795 $\pm$ 119 & 1.79 $\pm$ 0.01 & 7.14 $/$ 9\\
15 & 264.2 $\pm$ 0.5 & 6790 & 334 $\pm$ 33   &    \nodata   &     \nodata    & 5.82 $/$ 16 & 7344 $\pm$ 226 & 0.95 $\pm$ 0.01 & 10.63 $/$ 13\\
\hline
{Southwest}& {}              &  {}  & {}              &    {}        &     {}         & {}          & {}             & {}              & {} \\
16 & 281.5 $\pm$ 0.4 & 6122 & 1010 $\pm$ 171 & 303 $\pm$ 75 & 8891 $\pm$ 425 & 17.21 $/$ 15& 9169 $\pm$ 122 & 6.37 $\pm$ 0.01 & 7.88 $/$ 12\\
17 & 292.6 $\pm$ 0.7 & 7384 & 1206 $\pm$ 271 & 225 $\pm$ 78 & 8718 $\pm$ 339 & 3.39 $/$ 11 & 8089 $\pm$ 196 & 2.14 $\pm$ 0.02 & 1.18 $/$ 11\\
18 & 306.6 $\pm$ 0.5 & 7384 & 941  $\pm$ 274 & 261 $\pm$ 52 & 8718 $\pm$ 443 & 2.11 $/$ 11 & 8637 $\pm$ 253 & 3.70 $\pm$ 0.06 & 2.48 $/$ 11\\
19 & 338.0 $\pm$ 0.3 & 5796 & 351 $\pm$ 30   &    \nodata   &     \nodata    & 19.85 $/$ 19& 8145 $\pm$ 150 & 9.29 $\pm$ 0.04 & 5.87 $/$ 14\\
20 & 343.1 $\pm$ 1.8 & 7384 & 950 $\pm$ 72   &    \nodata   &     \nodata    & 6.21 $/$ 13 & 8169 $\pm$ 327 & 8.63 $\pm$ 0.32 & 3.80 $/$ 11\\
21 & 359.7 $\pm$ 0.3 & 5796 & 710 $\pm$ 63   & 153 $\pm$ 117& 9801 $\pm$ 442 & 11.89 $/$ 17& 8784 $\pm$ 125 & 7.56 $\pm$ 0.02 & 11.01 $/$ 14\\
\hline
{Northwest}& {}              &  {}  & {}              &    {}        &     {}         & {}          & {}             & {}              & {} \\
22 &  15.1 $\pm$ 0.3 & 6991 & 484 $\pm$ 37   &    \nodata   &     \nodata    & 6.36 $/$ 15 & 8914 $\pm$ 142 & 6.08 $\pm$ 0.02 & 0.83 $/$ 11\\
23 & 23.7 $\pm$ 0.5  & 6014 & 959 $\pm$ 130  & 102 $\pm$ 39 & 8183 $\pm$ 245 & 2.53 $/$ 16 & 8846 $\pm$ 191 & 6.09 $\pm$ 0.03 & 1.41 $/$ 11\\
24 & 47.3 $\pm$ 0.4  & 6014 & 289 $\pm$ 28   &    \nodata   &     \nodata    & 6.00 $/$ 18 & 8409 $\pm$ 156 & 5.29 $\pm$ 0.01 & 9.06 $/$ 13\\
25 & 67.8 $\pm$ 0.5  & 6014 & 544 $\pm$ 193  & 124 $\pm$ 25 & 7226 $\pm$ 440 & 1.82 $/$ 16 & 7896 $\pm$ 204 & 3.08 $\pm$ 0.02 & 6.92 $/$ 13\\
26 & 86.1 $\pm$ 0.4  & 5401 & 806 $\pm$ 155  & -138 $\pm$ 82& 7646 $\pm$ 270 & 11.69 $/$ 16& 7207 $\pm$ 72  & 7.94 $\pm$ 0.01 & 9.01 $/$ 15\\
{} &        {}       &   {} & 838 $\pm$ 187\tablenotemark{c}& {} &10931 $\pm$ 294\tablenotemark{c}& {} &{} &      {}         &     {}   \\
\enddata
\tablecomments{The horizontal lines denote the quadrants. The first column gives the spot numbers defined in Figure \ref{fig:spots}. The position angles in the second column are computed as the time averages of all observations from 2003 and onward and their uncertainties represent the standard deviations of the samples. Column 3 gives the starting time $t_{\rm start}$ which represents the time each spot begins to exhibit a systematic increase in its radial distance from the center of SN 1987A. Columns 4 to 7 correspond to the fitting of the radial distances. The variables  $v_{1}$ and $v_{2} $ correspond to the inclination-corrected radial expansion of the hotspots, while $t_{\rm break}$ represents the time of velocity change. The velocity and break time uncertainties are estimated from the covariance matrix. Column 7 gives the reduced chi-square and the degrees of freedom of the fitting of the radial distances. The last 3 columns (8-10) correspond to the flux fitting of the spots. The variable $t_{\rm peak}$ corresponds to the time of maximum brightness while $\mathrm{Flux_max}$ is the peak brightness. The last column gives the reduced chi-square and the degrees of freedom of the flux fitting. The uncertainties of the $t_{\rm peak}$ and $\rm Flux_{\rm max}$ are computed from Monte Carlo simulations.}
\tablenotetext{a}{For velocity fitting.}
\tablenotetext{b}{For flux fitting.}
\tablenotetext{c}{Third velocity/ second break time.}
\end{deluxetable*}

The best-fit models are shown in Figure \ref{fig:radial evolution}. Individual fits and residual plots are included in Appendix \ref{Appendix: Radial Fitting of the Spots}. Nine spots are best described by a first-order polynomial, sixteen spots by a piecewise function with one breakpoint, and only one spot (spot 26) by a piecewise function with two breakpoints. Table \ref{tab:velocities} summarizes the best-fit parameters and their uncertainties for each spot.

\begin{figure*}[htb!]
\plottwo{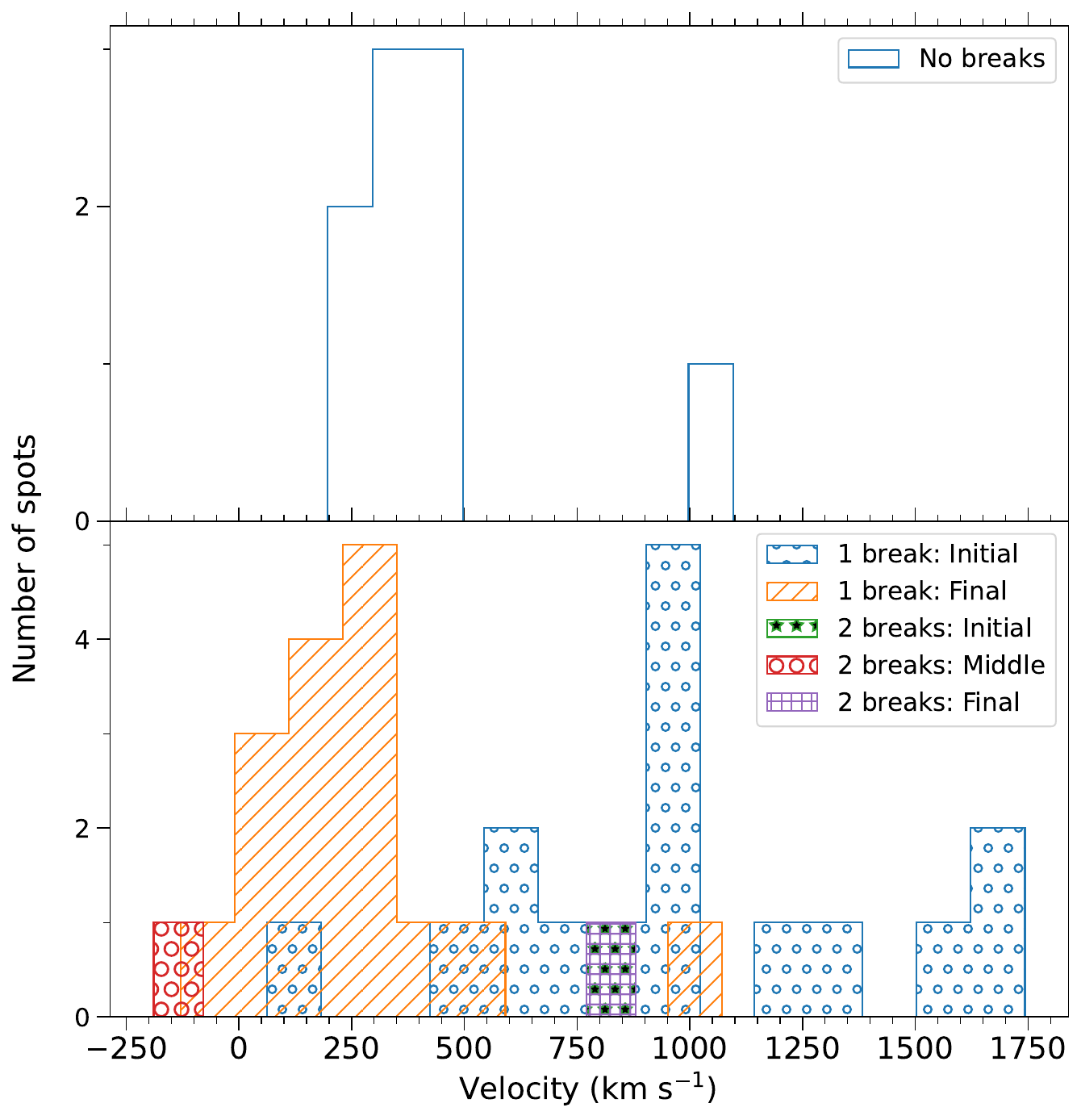}{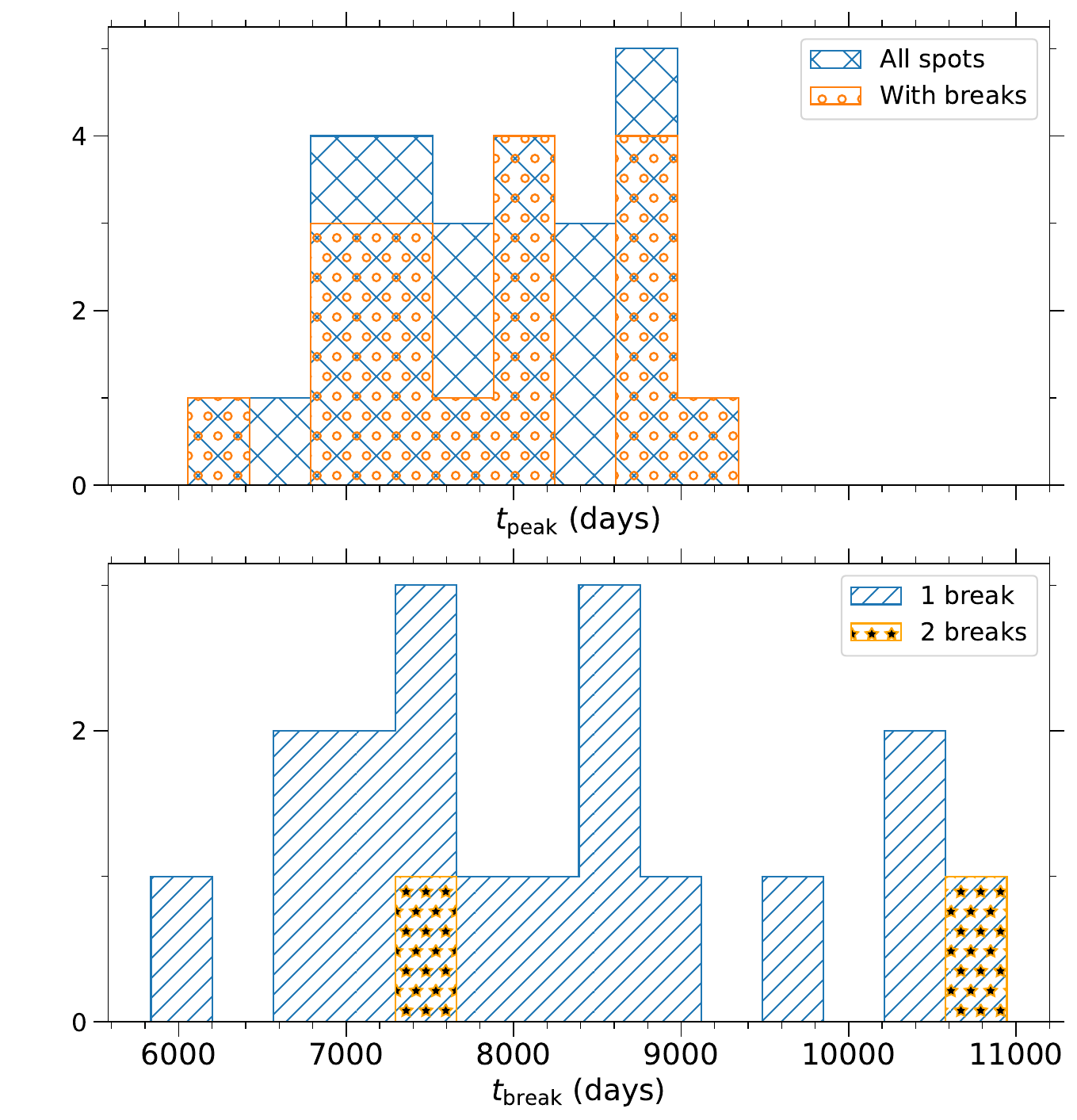}
\caption{Left panels: Velocity distributions of the spots based on their models: straight and piecewise lines with breaks. The initial velocities are plotted separately from the subsequent ones to showcase that the spots are slowing down as time passes. Right upper panel: Histograms of the flux peak times, estimated by the exponentially modified Gaussians. The histogram includes both the flux peak distribution of all spots and those that change velocity at least once. Right lower panel: The total distribution of all break times.\label{fig:velocity histogram}}
\end{figure*}

For cases with an absent break time, the 90th percentile range of velocities spans from $240$ to $\sim 830 \ \rm{km\ s^{-1}}$. For spots with one break time, the 90th percentile range of initial velocities ranges widely, from $390$ to $1660 \ \rm{km\ s^{-1}}$, with a predominant concentration around $\sim1000 \ \rm{km\ s^{-1}}$. Post break time, spots decelerate to velocities around $260 \ \rm{km\ s^{-1}}$, except for spot 7, which displays an acceleration. This acceleration corresponds to a phase when the spot's brightness decreases and becomes comparable to its surrounding background. Spot 26 was a special case with two velocity changes: an initial velocity of $806 \ \rm{km\ s^{-1}}$, followed by a decline to $-138 \ \rm{km\ s^{-1}}$, and ultimately reaching $838 \ \rm{km\ s^{-1}}$. We note, though, that all estimated negative velocities (spots 8 and 26) have large relative uncertainties and are not interpreted as a physical movement. The histograms in Figure \ref{fig:velocity histogram} showcase the velocity distributions for the different cases.

\begin{figure}[htb!]
\plotone{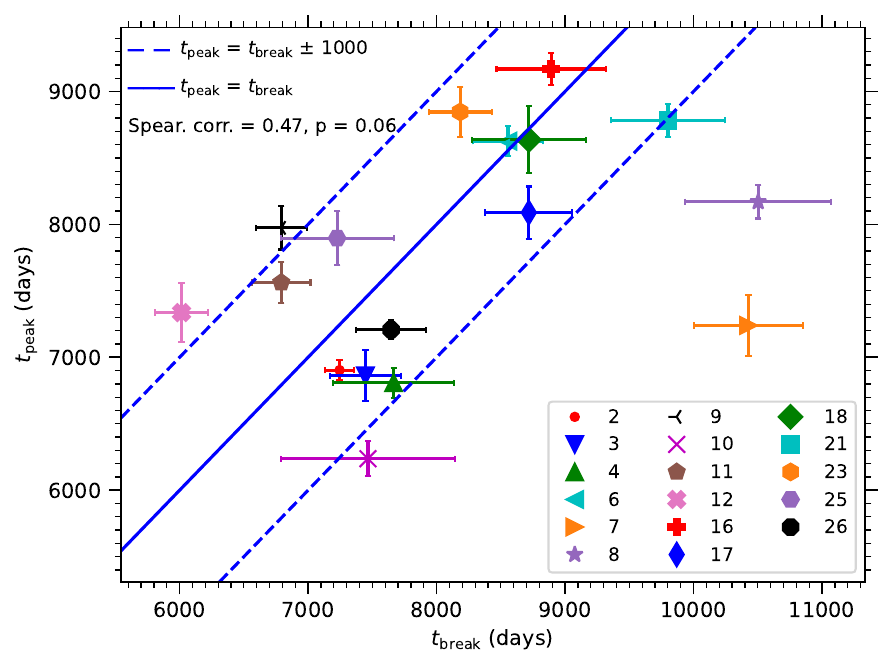}
\caption{Scatter plot demonstrating the relationship between the time of maximum flux and the instances when the hotspots undergo velocity changes. The dataset exhibits a Spearman correlation coefficient of 0.47 (p = 0.06), indicating a moderate positive correlation between the variables with marginal statistical significance.\label{fig:scatterplot breaks}}
\end{figure}

\begin{figure}[htb!]
\plotone{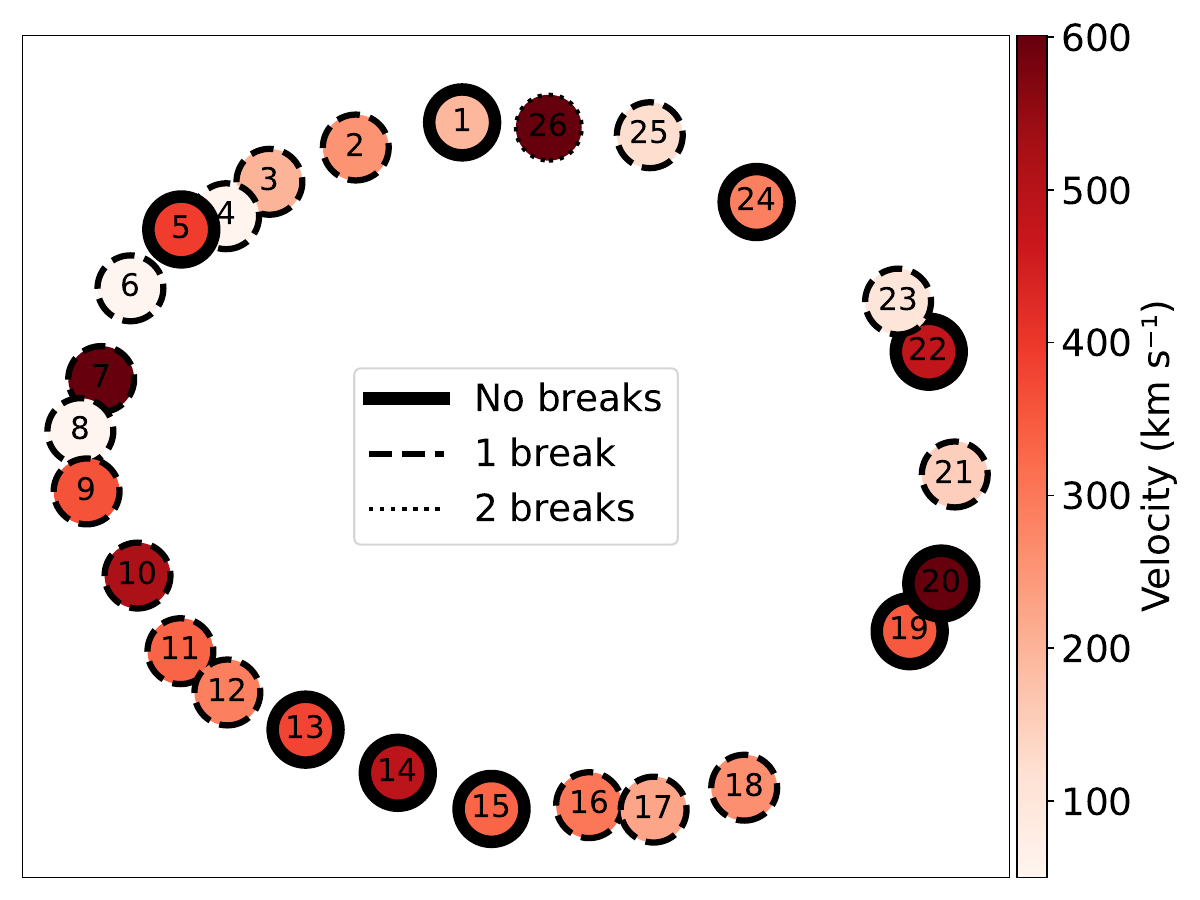}
\caption{Illustration of the number of breaks and velocities of the spots across the ER. The spots are color-coded with their velocities, with the final velocity shown for cases where a velocity change occurred.\label{fig:vmap}}
\end{figure}

The histograms in the right panel of Figure \ref{fig:velocity histogram} depict the distribution of $t_{\rm break}$ and $t_{\rm peak}$. The 90th percentile range of $t_{\rm break}$ spans from 6700 to 10,600 days, with a median of $\sim 7900$ days. Similarly, the distribution of $t_{\rm peak}$ mirrors that of the break times, ranging from 6500 to approximately 9000 days, with a median of $\sim 7800$ days. The scatter plot in Figure \ref{fig:scatterplot breaks} implies a monotonic relationship between these two parameters. We quantified this relationship with Spearman correlation analysis, which yielded a moderate positive correlation coefficient of 0.47 (p-value = 0.06). Generally, flux peaks appear to occur within a range of $\pm 1000$ days from the break times for each spot. There is a slight predominance of spots where break times occur after their flux peaks during the gradual fading phase. Figure \ref{fig:vmap} visualizes how the number of break times and velocities of the spots are distributed across the ER,  with final velocities depicted for cases where a break time occurs. The results reveal an even occurrence of different models (with or without breaks) throughout the entire ER. Spots in the southeast typically have higher final velocities, while the lowest velocities are observed in the northeast. 

The correlations among all the involved parameters in Table \ref{tab:velocities} were also examined to identify potential patterns. We observe that hotspots emerging later tend to exhibit delayed velocity changes and reach their maximum brightness at later times. Specifically, a moderate positive correlation coefficient of 0.57 (p-value = 0.002) was found between $t_{\rm peak}$ and $t_{\rm start}$, while a similar correlation coefficient of 0.55 (p-value = 0.02) was observed between $t_{\rm start}$ and $t_{\rm break}$. Additionally, western spots generally display larger $t_{\rm peak}$, $t_{\rm start}$, and $\rm Flux_{\rm max}$ compared to their eastern counterparts. The earliest break times are observed by spots in the southeast (spots 9-12), followed by some in the northeast (spots 2-4) and those in the northwest (spots 23, 25, 26), most of them typically occurring before 8000 days. Notably, spots in the southeast have the lowest fluxes (see Figure \ref{fig:radial evolution}).

\subsection{Sizes, Elongation and Orientation}\label{subsection: Sizes}

\begin{figure}
\plotone{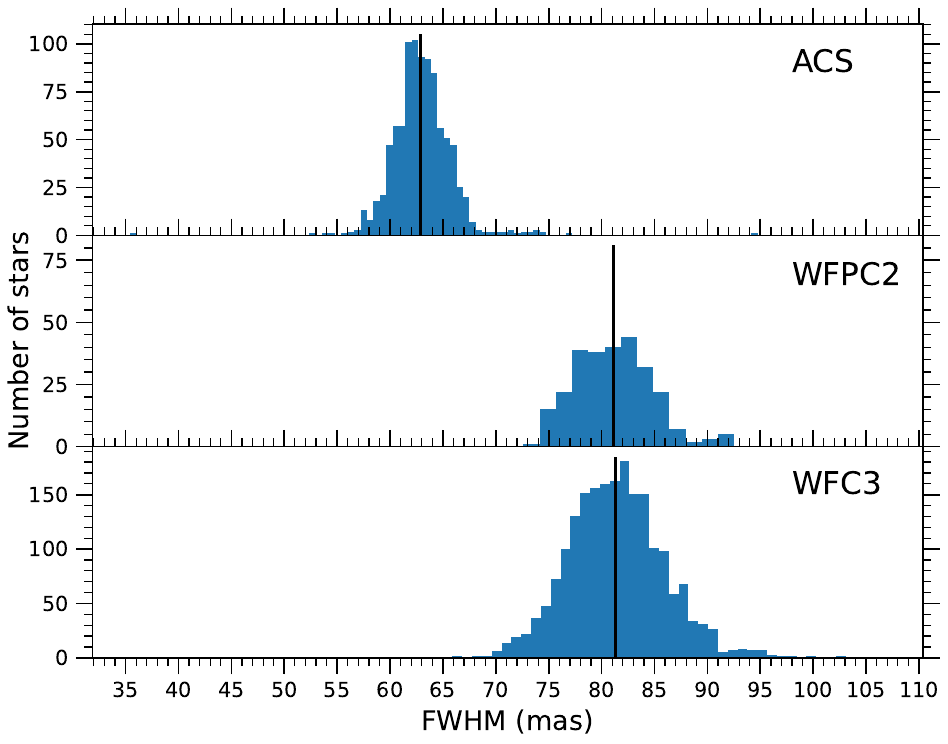}
\caption{The FWHM distribution of stars near SN 1987A, fitted by the 2D Gaussian model (Appendix \ref{Appendix: Fitting}). The distributions are categorized based on the instruments used (ACS, WFPC2, and WFC3) and encompass the FWHM of stars fitted across all epochs. The major and minor FWHM were treated as independent in the analysis.\label{fig:star sizes}}
\end{figure}

To determine if the spots are spatially resolved, we compare their fitted FWHM to the spatial resolution of the instruments. Fitting functions to the pixelated PSF yield larger FWHM values than the instrumental PSF FWHM \citep{WFC3}. For that reason, we determined the spatial resolution by fitting stars in the images using the same Gaussian model employed for the hotspots. The \texttt{photutils} \texttt{DAOStarFinder} tool was used to identify the brightest stars with a high signal-to-noise ratio, located away from the edges of the images to avoid undesirable FOV effects \citep{Acs, WFC3}. Since the resolution is influenced by the instrument and dither pattern, we compiled three distinct star samples categorized by the instruments (WFCP2, ACS, and WFC3). Each sample consisted of stars with a signal-to-noise ratio above 10, selected from observations spanning from day 5796 to day 12,598. We observed minor temporal variations of the mean FWHM at the 4\% level for WFC3 and the 1\% level for the other instruments, as expected. However, these variations do not significantly impact the results, allowing us to assume a constant resolution for each instrument over the years. The calculated widths of the major and minor axes of the fitted stars are consistent, with elongation effects typically below 8\%. The combined samples are depicted in Figure \ref{fig:star sizes}. The mean resolution for each instrument is determined by taking the mean FWHM of the respective sample (Table \ref{tab:instrument resolution}). 

We consider a spot resolved if it satisfies the following condition:
\begin{equation}\label{eq:cut-off}
\rm{FWHM_{spot, obs}}-3\sigma_{\rm{FWHM_{spot, obs}}} > S_{Cut-off}, 
\end{equation}
where $S_{\rm{Cut-off}}$ is the resolution plus 3 standard deviations, see Table \ref{tab:instrument resolution}, and $\sigma_{\rm{FWHM_{spot, obs}}}$ is the uncertainty of the FWHM of each spot. Three standard errors are subtracted to reduce the probability of falsely concluding that a spot is resolved due to the uncertainties of the estimated FWHM. For simplicity, given that each spot is characterized by both a major and a minor FWHM, we used the circularized FWHM as the $\rm{FWHM_{spot, obs}}$, calculated as $\left ( 0.5 \left ( \mathrm{FWHM}^{2}_{\mathrm{minor}} + \mathrm{FWHM}^{2}_{\mathrm{major}} \right ) \right )^{1/2}$. Thirteen out of the twenty-six hotspots were identified as resolved. Some hotspots met the resolution criteria only during specific observations, while others remained unresolved throughout all epochs. We classify a hotspot as resolved if it meets the resolution condition in at least three observations.

\begin{deluxetable}{lhhchhc}[htb]
\tabletypesize{\scriptsize}
\tablecaption{Mean FWHM of stars for every instrument \label{tab:instrument resolution}}
\tablehead{
\colhead{Instrument} & \colhead{} & \colhead{} & \colhead{Mean FWHM} & \colhead{} & \colhead{} & \colhead{Cut-Off \tablenotemark{b}} \\
\colhead{} & \colhead{} & \colhead{} & \colhead{(mas)} & \colhead{} & \colhead{} & \colhead{(mas)}
} 
\startdata
ACS & & & 62.88 & & & 69.62 \\
WFPC2\tablenotemark{a} & & & 81.13 & & & 92.00 \\
WFC3 & & & 81.28 & & & 94.24 \\
\enddata
\tablenotetext{a}{For observations between 2007 and 2009.}
\tablenotetext{b}{Mean FWHM plus 3 Standard deviations of the star samples (Figure \ref{fig:star sizes}). This is used as a limit when determining if the hotspots are resolved. }
\end{deluxetable}

A correction formula was applied to account for the instrumental broadening of the spots:
\begin{equation}
\label{eq:spot size}
    \rm{FWHM_{spot}}= \sqrt{{\rm{FWHM}^{2}_{spot, obs}}-{FWHM^{2}_{star}}} ,
\end{equation}
where $\rm{FWHM_{spot, obs}}$ is the measured circularized FWHM and $\rm{FWHM_{star}}$ the size of the unresolved stars according to Table \ref{tab:instrument resolution}. Equation \ref{eq:spot size} is valid as long as a spot is larger than the estimated FWHM of the stars in each observation. This condition is always met for the resolved spots due to Equation \ref{eq:cut-off}. In contrast, no correction was applied for spots that did not meet the resolution condition. In such cases, an estimated upper limit for their sizes was assigned ($6.03 \times 10^{16} \ \rm cm$), derived from the mean FWHM of stars in WFC3 observations.

\begin{figure*}[tbh!]
\plotone{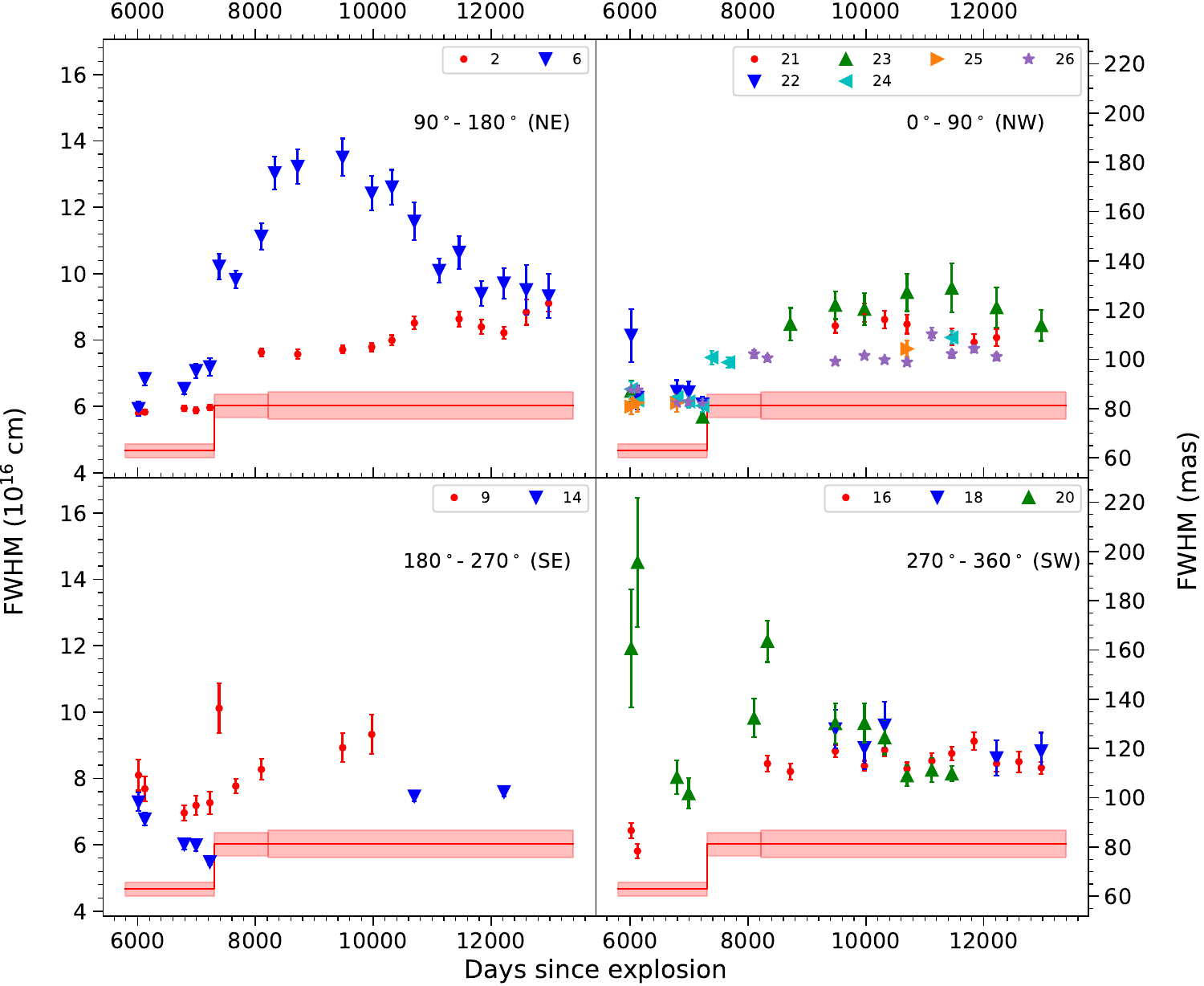}
\caption{Time evolution of the observed circularized FWHM of all the resolved spots. The error bars here are 3 standard errors. The red solid line and the shaded region are the average resolution and 3 standard deviations measured from the star samples for the different instruments. The left is the corresponding physical distances in cm, while the right y-axis is in mas.\label{fig:spot size}}
\end{figure*}

The temporal evolution of the observed circularized FWHM of the resolved spots is shown in Figure \ref{fig:spot size}. The systematic spike observed during the transition from the ACS to the WFPC2 observations in the uncorrected sizes is smoothed out after correcting for instrumental broadening. There is no evidence of a systematic increase in size for the resolved hotspots over time in any direction, with the only exceptions being hotspots 2 and 6. In half of the resolved hotspots, the minor FWHM is consistent with the defined cut-off level, while the major FWHM is always well above the resolution limit. We note that spots with the highest fluxes are generally resolved, except for spot 19, the brightest one.

We computed a time average of the corrected circularized FWHM for the resolved hotspots (Equation \ref{eq:spot size}), considering only the observations that met the resolution condition. The results are reported in Table \ref{tab:spot size}. For unresolved hotspots, we report the mean FWHM of stars in the WFC3 observations (from Table \ref{tab:instrument resolution}) as an upper limit. The results indicate a circularized FWHM corresponding to physical sizes between $4 \times 10^{16} \ \rm{cm}$ and $6 \times 10^{16} \ \rm{cm}$ for the majority of hotspots.

The time averages of the angular orientation ($\phi$) and axis ratio of the hotspots are also provided in Table \ref{tab:spot size}. The orientation $\phi$ is defined as the angle of the major axis of the hotspots relative to their radial direction (PA), measured counterclockwise. Notably, many hotspots exhibit elongation. For the majority of the hotspots, there is no systematic time evolution of the elongation. As anticipated, unresolved spots tend to be more symmetric than resolved ones. As a reference, the widths of the fitted stars in the observations had axis ratios between 0.89 and 0.95, with the more elongated cases in the WFC3 observations. All spots deemed resolved had ratios less than these, while most of the unresolved ones closely approached these values. Nevertheless, instances exist where the elongation for unresolved spots falls below 0.8 (spots 1, 5, 8, 15). This occurs because the major widths exceed the resolution cut-off while the minor widths do not. Consequently, the circular widths end up below the cut-off, resulting in the classification of these spots as unresolved. In terms of orientation, hotspots generally maintain a constant orientation over time, with only a few exceptions (spots 3, 16, 17, 19, 20). Overall, most of the spots appear to be elongated in a direction almost perpendicular to their PA.

\begin{figure}[tbh!]
\plotone{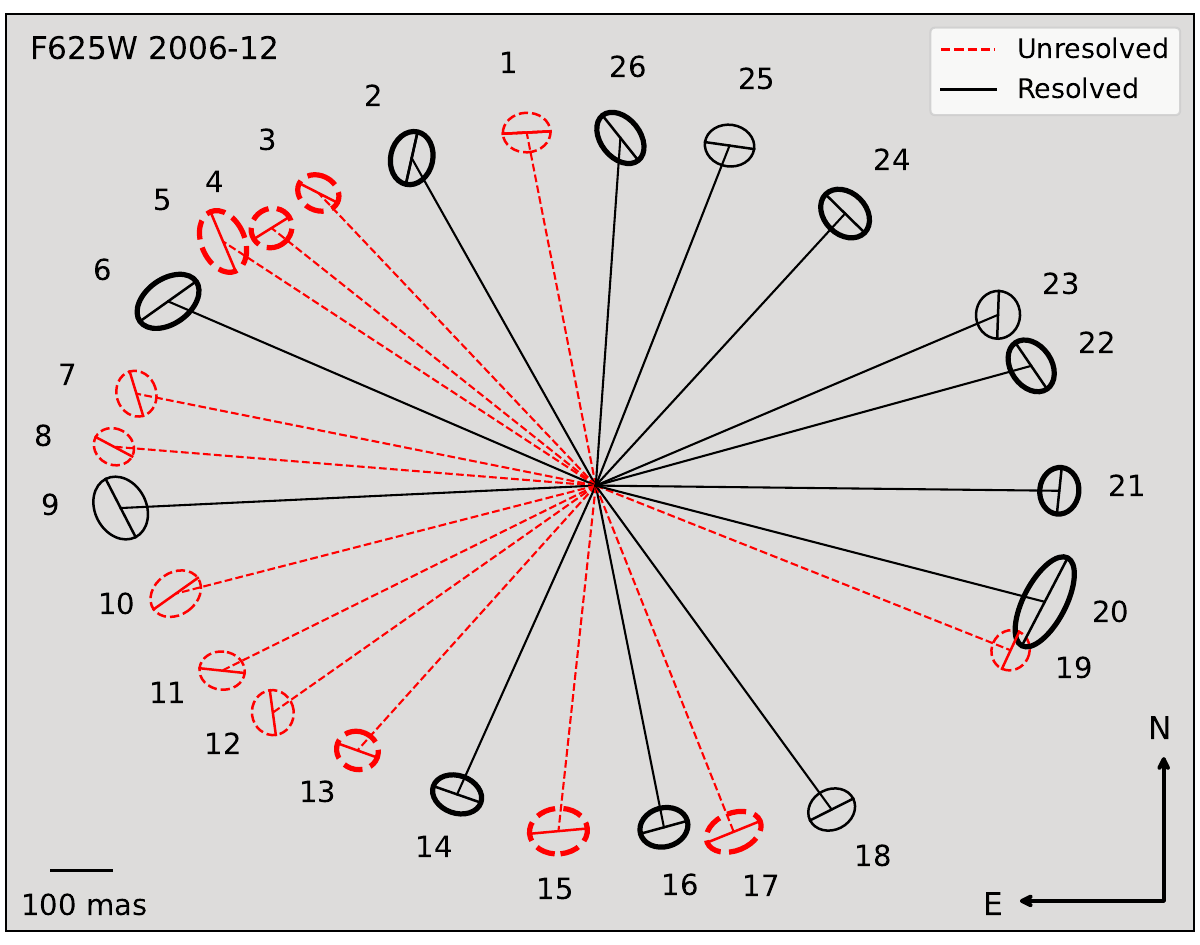}
\caption{Illustration of the shapes and the orientation of the spots relative to their position angle (PA) in the ACS/F625W image from 7226 days (2006 December 6). The sizes of the ellipses represent the FWHM of the spots, while the lines originating from the center represent their PA. The bold ellipses in the figure represent hotspots with an orientation uncertainty less than $7.5^{\circ}$, where the uncertainty is one standard deviation of the combined statistical error. The red dashed lines are the unresolved spots and the black solid lines are the resolved spots. The classification of the resolved spots is based on the resolution condition in Equation \ref{eq:cut-off} on the whole dataset. Notably, spots 21 and 18 fulfill the resolution condition in the late WFC3 observation. \label{fig:2006 orientation}}
\end{figure}

The orientation and elongation of the spots are illustrated in Figure \ref{fig:2006 orientation}. The plot showcases the results from the ACS/F625W observation taken 7226 days after the explosion. Notably, spots 5 and 15 are unresolved despite their larger size compared to other resolved spots. The reason for this is the large uncertainties on the FWHM, which are taken into consideration in the resolution condition in Equation \ref{eq:cut-off}. The significant uncertainties arise from the relatively faint nature of these spots and their location in a very crowded region. A similar case is spot 18, which is considered resolved in the WFC3 observations but not in the ACS ones due to the substantial width uncertainties.

\begin{deluxetable*}{cccc cccc}[tbh!]
\tabletypesize{\scriptsize}
\tablewidth{0pt} 
\tablecaption{FWHM, axis ratios and orientations of the spots\label{tab:spot size}}
\tablehead{
\colhead{Spot} & \colhead{Circularized FWHM} & \colhead{Axis ratio} & \colhead{$\phi$} & \colhead{Spot} & \colhead{Circularized FWHM} & \colhead{Axis ratio} & \colhead{$\phi$}\\
\colhead{} & \colhead{($10^{16}\ \rm{cm} \equiv 13.48$ mas)} & \colhead{} & \colhead{(deg)} & \colhead{} & \colhead{($10^{16}\ \rm{cm} \equiv 13.48$ mas)} & \colhead{} & \colhead{(deg)}
}
\startdata
{Northeast}& {}              & {}              & {}             & {Southwest}& {}              & {}                  & {}\\ 
1 & $<$ 6.03        & 0.77 $\pm$ 0.05 & 82.6 $\pm$ 1.5 & 16& 5.79 $\pm$ 0.88 & 0.67 $\pm$ 0.07 & 84.9 $\pm$ 0.9\\ 
2 & 4.42 $\pm$ 1.07 & 0.79 $\pm$ 0.03 & 140.9 $\pm$ 0.8& 17&  $<$ 6.03       & 0.81 $\pm$ 0.10 & 117.7 $\pm$ 3.9\\
3 & $<$ 6.03        & 0.89 $\pm$ 0.06 & 19.0 $\pm$ 7.5 & 18& 6.83 $\pm$ 0.53 & 0.62 $\pm$ 0.10 & 77.3 $\pm$ 1.9\\
4 & $<$ 6.03        & 0.94 $\pm$ 0.03 &106.6 $\pm$ 10.3& 19& $<$ 6.03        & 0.90 $\pm$ 0.04 & 88.6 $\pm$ 3.1\\
5 & $<$ 6.03        & 0.69 $\pm$ 0.07 & 144.5 $\pm$ 1.6& 20& 6.47 $\pm$ 1.28 & 0.61 $\pm$ 0.11 & 97.2 $\pm$ 1.5\\
6 & 7.05 $\pm$ 2.52 & 0.59 $\pm$ 0.07 & 60.7 $\pm$ 0.9 & 21& 5.85 $\pm$ 0.44 & 0.74 $\pm$ 0.07 & 79.7 $\pm$ 7.2\\
\cline{5-8}
7 & $<$ 6.03        & 0.83 $\pm$ 0.07 & 121.7 $\pm$ 2.0& {Northwest}& {}              & {}                  & {}\\ 
8 & $<$ 6.03        & 0.77 $\pm$ 0.05 & 149.2 $\pm$ 1.6& 22& 4.36 $\pm$ 0.64 & 0.81 $\pm$ 0.07 & 97.1 $\pm$ 5.9\\
\cline{1-4}
{Southeast}& {}              & {}              & {}& 23& 5.14 $\pm$ 1.60 & 0.74 $\pm$ 0.12 & 71.8 $\pm$ 3.7\\
9 & 5.77 $\pm$ 0.72 & 0.74 $\pm$ 0.06 & 114.3 $\pm$ 2.1& 24& 4.28 $\pm$ 0.42 & 0.79 $\pm$ 0.03 & 84.9 $\pm$ 1.3\\
10& $<$ 6.03        & 0.84 $\pm$ 0.09 & 36.0 $\pm$ 17.7& 25& 4.08 $\pm$ 0.39 & 0.78 $\pm$ 0.04 & 95.6 $\pm$ 1.5\\
11&  $<$ 6.03       & 0.86 $\pm$ 0.07 & 142.5 $\pm$ 3.8& 26& 4.36 $\pm$ 0.36 & 0.71 $\pm$ 0.03 & 43.2 $\pm$ 0.5\\
12& $<$ 6.03        & 0.86 $\pm$ 0.08 & 96.1 $\pm$ 7.3  \\
13& $<$ 6.03        & 0.90 $\pm$ 0.03 & 101.9 $\pm$ 1.6 \\
14& 4.23 $\pm$ 0.87 & 0.78 $\pm$ 0.06 & 95.7 $\pm$ 0.9  \\
15& $<$ 6.03        & 0.83 $\pm$ 0.08 & 103.3 $\pm$ 3.7 \\
\enddata
\tablecomments{The horizontal lines denote the quadrants. The circularized FWHM values represent weighted time averages, computed exclusively for spots that meet the resolution condition in at least three observations. The upper limit of $6.03 \times 10^{16}\ \rm{cm}$ corresponds to the mean FWHM of the stars in the WFC3 observations. For comparison, the sizes of the stars and resolution cut-off in the ACS observations reported in Table \ref{tab:instrument resolution} correspond to $4.67 \times 10^{16}\ \rm{cm}$ and $5.17 \times 10^{16}\ \rm{cm}$, respectively. All FWHM uncertainties are one standard deviation of the samples. The axis ratio represents the weighted mean of all observations (ACS, late WFPC2, and WFC3). The axis ratio errors are standard deviations of the samples. The orientation ($\phi$) is defined as the angle of the major axis of the spots relative to their radial direction in the ER (PA), measured counterclockwise. The orientation values represent the averages of all observations while the error is measured by error propagating the individual uncertainties of each observation.}
\end{deluxetable*}

\section{Discussion}\label{sec: Discussion}

The ER of SN 1987A provides an excellent opportunity for studying shock interactions, particularly through the observation and analysis of the hotspots that emerged as a consequence of its collision with expanding debris. We follow the temporal evolution of these hotspots through regular monitoring with HST over nearly three decades. This extended observational timeline provides access to unique data, enabling us to trace the progression of shocks over time.

During the earliest epochs probed by our HST observations ($\sim 2800-5000$ days), the observed emission likely comprises a blend of diffuse radiation from the ER and emission from denser regions or substructures flash-ionized by the SN. As the outer ejecta collide with the ER and its denser obstacles, the emission budget becomes dominated by the higher-density regions situated closer to the center of SN 1987A. This scenario offers the most plausible explanation for the observed radial decline in several hotspots shortly after their detection (see Figure \ref{fig:spot 2 detection}). In particular, if the inner obstacles have higher densities ($ \gtrsim7.5\times 10^{4}\ \mathrm{cm^{-3}}$) they would have cooled and recombined before our first observations at $\sim 2800$ days, rendering them invisible in the early images.  Only upon being shocked by the ejecta would they become visible again.

We measure the radial expansion of the hotspots from the center of SN 1987A and correct for projection effects due to the inclination of the ER. Approximately two-thirds of the spots exhibit a break in the expansion rate, transitioning from a higher initial median velocity of $960 \ \rm km \ s^{-1}$ (with a 90th percentile range of $390-1660 \ \rm km \ s^{-1}$) to a lower median velocity of $260\ \rm km \ s^{-1}$ (with a 90th percentile range of $40-660 \ \rm km \ s^{-1}$). Since only spot 26 exhibited two velocity changes, we disfavor the possibility of an underlying physical mechanism and attribute this behavior to a combination of higher uncertainty in the fits and a complex background. Typically, the time of the velocity breaks aligns with the hotspots reaching their peak brightness, commonly observed between 7000 and 9000 days. Hotspots that maintain a constant velocity have a narrower velocity distribution with a median of $ 380\ \rm km \ s^{-1}$ (90th percentile range of 230-$830 \ \rm km \ s^{-1}$), similar to the final velocities of the spots that show a break. For comparison, \cite{Fransson_2015} reports velocities ranging from $180\ \rm{km\ s^{-1}}$ to $950\ \rm{km\ s^{-1}}$, with indications of a decreasing velocity after approximately 8000 days for some of the spots. These findings, however, were based on observations in the R-band and B-band between 1994 and 2014, using a one-dimensional analysis and a linear fit, without correction for the inclination of the ER. Despite these disparities, we observe similar patterns in the radial evolution of the same hotspots.

When interpreting the time evolution of velocities, it is crucial to note that the hotspots we observe represent the two-dimensional projections of three-dimensional emitting volumes of ER material through which the forward shock has passed. Consequently, the observed motion of the spots is influenced not only by the shock velocity but also by other factors such as the obstacle geometry and density, and cooling time of the shocked gas.

The cooling time $t_{\rm cool}$ denotes the duration for the post-shocked gas to thermally collapse and cool down to temperatures of order $ 10^{4} \ \rm K$ \citep{pun_2002, Groeningsson_2008}. At these temperatures, the cooling is balanced by photoionization heating. The resulting thermal equilibrium is responsible for the optical emission observed from the hotspots \citep{pun_2002}. When the cooling time is comparable to the timescale for the blast wave to traverse the obstacle, $t_{\rm cross}$, we are likely observing emission originating from layers of cooled gas across the entirety of the hotspot's sides. In this scenario, as time progresses, the optical emission increases as layers of gas with higher velocities across the whole spot undergo thermal cooling. In contrast, if the cooling time is significantly shorter than the characteristic $t_{\rm cross}$, we may observe emission only from a thin layer behind the shock. Under such conditions, observation at infinite resolution would reveal the projected partial volume of the spot, appearing flattened with an orientation dependent on the direction of the shock and the density profile of the spot. Figure \ref{fig:2006 orientation} and Table \ref{tab:spot size} reveal a consistent angular orientation of the hotspots within the image plane. Most hotspots exhibit elongation perpendicular to their PA, suggesting a short cooling time. One caveat to consider is the limited resolution of the instruments and the fact that only half of the spots are spatially resolved. This introduces uncertainty regarding the actual orientation of the hotspots, especially when combined with background diffuse emission.

If the temperature behind the shocks is below $2 \times 10^{7} \ \rm K$, cooling is primarily driven by line emission and the cooling time can be estimated by \citep{sn1987_review}:
\begin{equation}
    t_{\mathrm{cool}} = 38 \ \left (\frac{v_{\mathrm{shock}}}{500 \ \mathrm{km \ s^{-1}}}\right )^{3.4} \ \left ( \frac{n_{\mathrm{e}}}{10^{4} \ \mathrm{cm^{-3}}} \right )^{-1} \ \mathrm{years}
\end{equation}
where $n_{\rm e}$ is the pre-shock electron density and $v_{\rm shock}$ the velocity of the shock. The majority of the hotspots become visible between 5000 and 6000 days, suggesting an upper limit for the cooling time of approximately 14 years. Flash-ionization modeling has found preshock densities in the ER of up to $\sim 3\times10^4$ cm$^{-3}$ \citep{Lundqvist_1996, Mattila_2010}. Based on these densities, the fastest shocks capable of producing optical emission, consistent with our cooling time, are around $\sim 500 \ \rm km \ s^{-1}$. Since the cooling time significantly depends on the velocity of the transmitted shock, faster shocks would require even higher densities to match our expected cooling time. For instance, a shock velocity of $\sim 1000 \ \rm km \ s^{-1} $, comparable to the high initial velocities observed from spots with a break, would require densities of $\geq 4\times 10^{5} \ \rm cm^{-3}$. As we will discuss below, such high densities cannot be justified. 

We can estimate the velocity of a shock driven by the expanding ejecta into the dense clumps by \citep{ Sgro_1975}:
\begin{equation}
\label{eq: v_shock}
v_{\mathrm{shock}} = \left ( \frac{\rho_{\mathrm{H II}}}{\rho_{\mathrm{clump}}} 
 \delta  \right )^{1/2} \ v_{\mathrm{blast}},
\end{equation}
where $v_{\rm{blast}}$ is the velocity of the debris in the H II region located inside the ER (e.g., \citealt{Chevalier_1995}) with density $\rho_{\rm{H II}}$, $\rho_{\rm{clump}}$ is the density of the obstacle and $\delta$ is a function that depends on the obliquity of the shock and the preshock density ratio $\rho_{\mathrm{clump}} /  \rho_{\mathrm{H II}}$. The quantity $\delta$ is maximized for head-on collisions, such as those occurring at the tip of the protrusion, and rises monotonically from 1 to 6 as the preshock density increases ($\rho_{\mathrm{clump}} /  \rho_{\mathrm{H II}} \rightarrow \infty $, $\delta \rightarrow 6$), with a value of $\sim 4$ at a preshock density ratio of 100 \citep{Sgro_1975}. More details about the range of $\delta$ can be found in Figure 9 in \citep{pun_2002}. By modeling the observed radio emission, \cite{Manchester_2002} measured the $v_{\rm{blast}}$ to be $\sim 3500\ \rm km \ s^{-1}$ with a 30\% uncertainty. Additionally, \cite{Borkowski_1997}, based on observations in X-ray wavelengths before the emergence of the first hotspot, estimated a $v_{\rm{blast}}$ of around $4100\ \rm km \ s^{-1}$ and a density $\rho_{\rm{H II}}$ of $150\ \rm cm^{-3}$. More recent estimates from X-rays give even higher blast wave velocity, $\sim 6700\ \rm km \ s^{-1}$ \citep{Ravi_2024}. 

Based on the above and with $\delta < 6$ to account for obliquity, Equation \ref{eq: v_shock} yields shock velocities of $\lesssim 1100 \ \rm km \ s^{-1}$ for a clump with a preshock density of $3\times 10^4$ cm$^{-3}$. From these shocks though, only those with velocities $\lesssim 500 \ \mathrm{km\ s^{-1}}$ would have time to cool and become visible in our observations. Such low velocities are in agreement with our estimated lower peak velocity from spots with no break and final velocities of spots with one break (see Figure \ref{fig:velocity histogram}). In contrast, shocks with higher velocities, such as the most extreme velocities of $\gtrsim 1000\ \mathrm{km\ s^{-1}}$ (Figure \ref{fig:velocity histogram}), would imply a cooling time of order 100 years and should only be traceable in the X-ray band.

Notably, the low-velocity peak in Figure \ref{fig:velocity histogram} is comparable to Doppler shifts of H lines in spectra of the ER, where typical velocities of $\sim 200 \ \rm km \ s^{-1}$ and FWHM values of a few hundred $\rm km \ s^{-1}$ are observed \citep{Groeningsson_2008, Jones_2023}. This agreement further suggests that these velocities indeed reflect the shock velocities transmitted through the ER material.

We investigate the possibility of systematic uncertainty in the early velocities, stemming from the fact that this is measured from our defined activation times when the radial distances reach a minimum. The hotspots start appearing before these times, but their emission is weak relative to the outer diffuse component, leading to larger inferred radial distances (illustrated in  Figure \ref{fig:spot 2 detection}). To obtain a limit on the maximal impact of this, we extend our analysis by incorporating an additional data point for hotspot 2 at 2932 days, coinciding with the observed radial recession (Figure \ref{fig:spot 2 detection}). We consider the scenario that the radial distance of the spot measured at this time is biased by the diffuse outer component and, therefore, replace it with the radial distance at 4336 days, the activation time. Due to the absence of information regarding the potential uncertainty associated with this additional point, we conducted the fitting without applying any weights. The resulting velocity is reduced by approximately 12\% (to $\sim 1500 \ \rm km \ s^{-1}$). Applying a similar reduction to the other high velocities would still yield values exceeding $900 \ \rm km \ s^{-1}$, which is higher than the expected velocities of radiative shocks. 

To resolve this discrepancy, we propose that the initial high velocities we observe are phase velocities resulting from density inhomogeneities within the clumps, coupled with relatively short cooling times. Specifically, we view each hotspot as an extended region comprising several small, dense substructures distributed within a less dense diffuse region. As the blast wave collides with the ER and its obstacles, the emission is primarily dominated by the higher-density regions closest to the center of the explosion. Subsequently, as the shock fronts advance within the low-density regions, the volumetric emission shifts toward the newly shocked dense substructures located further out, leading to a change in the radial position of the centroids. In this case, we observe a combination of the movement of the shocked substructures and the rate at which these outer substructures are being shocked. The latter depends on the propagation speed of the shocks through the less dense component, which should be higher than the shock velocity in the denser regions. 

Our proposed scenario gives a natural explanation for the early phase velocities. These velocities are intermediate between the low spot velocities after the breaks and the velocity of the blast wave in the H II region. They are also similar to the X-ray expansion rate of the ER of $1830\ \mathrm{km \ s^{-1}}$ \citep{Ravi_2024}. \cite{Mattila_2010} found low-density structures within the ER of $10^{-3} \ \mathrm{cm^{-3}}$ by fitting line fluxes from 300-2000 days with a photoionization model. A low-density component within the clumps of that order is consistent with the magnitude of the early phase velocities. Therefore, only subsequent final velocities are indicative of actual radiative shock velocities. It is worth noting that several spots fitted with a single velocity exhibited fluctuating radial behavior during early epochs (Figure \ref{fig:radial evolution} and Appendix \ref{Appendix: Radial Fitting of the Spots}). This behavior likely stems from the same underlying mechanism as the initial phase velocity observed in spots with a break time. We report combined velocities that we consider to correspond to actual shock velocities (the ``No breaks" and ``1 break: Final" distributions in Figure \ref{fig:velocity histogram}) with a 90th percentile range from 60 to $900 \ \rm km \ s^{-1}$ and a median of $300 \ \rm km \ s^{-1}$.

The early phase velocity terminates when the entire hotspot is shocked, leading to emission originating from all its substructures. This transition manifests as a break in the expansion rate of the hotspots. Furthermore, if the optical emission originates from substructures within the obstacles with short cooling times, we anticipate the maximum brightness to occur when the total volume of the clumps is shocked. This is supported by the light curves of the hotspot, which demonstrate a decay temporally correlated with the break times (Figure \ref{fig:radial evolution} and \ref{fig:scatterplot breaks}). Most hotspots exhibit a flux decrease within $\pm 3$ years after their change in the velocity slope and predominantly between 7000 and 9000 days. This is consistent with the timescale of the blast wave exiting the ER \citep{Fransson_2015}. Specifically, \cite{Fransson_2015} and \cite{Larsson_2019} report increased emission and the appearance of new faint hotspots outside the southeast part of the ER beyond 9500 days. We also found that the earliest velocity changes and the highest final velocities occur in the southeast quadrant (see Figure \ref{fig:vmap}). Assuming these velocities reflect the actual radiative shock velocities, they could be due to a lower density in this part of the ER. 

Some of the delayed velocity breaks that occur after 10,000 days (see Figure \ref{fig:scatterplot breaks}) could be attributed to the contribution of newly shocked gas as the blast wave extends to the outer regions of the ER, along with the gradual dissipation of the spots. The fitting routine of the spots, being two-dimensional, could be affected by such emission, resulting in abrupt breaks in the slope. An indicative instance of this scenario is observed in spot 7, the only spot that accelerates after its break. Spot 7 starts to fade after 8000 days as emission from the gas farther out becomes increasingly dominant within the fitting region. Related to this, we note that the new spots that appear in the southeast part of the ER, reported in \cite{Larsson_2019} and \cite{Fransson_2015}, are so faint and far away from the ER that they do not impact our fits in that quadrant (spots 10-13).

The measured sizes and fluxes of the hotspots can be used to estimate their filling factor $f$, defined as the fraction of hotspot volume filled by optically emitting high-density material. For simplicity, we will be assuming clumps with a spherical geometry, though an ellipsoid geometry with an axis ratio of $\sim 0.75$ (as observed for many of the spots, Table \ref{tab:spot size}) would yield similar results. We note that the density of the shocked gas, as indicated by nebular diagnostics based on several narrow emission lines in the optical and UV, ranges from $10^{6}$ to $10^{7}\ \rm{cm}^{-3}$ \citep{pun_2002, Groeningsson_2008_02}. A filling factor of 1 and a uniform density of $\sim 10^{6} \ \rm cm^{-3}$ would yield a total mass for the ER of the order of $\sim 1 \ M_{\odot}$ based on our measured sizes of the hotspots (Table \ref{tab:spot size}). For comparison, \cite{Lundqvist_1996} and \cite{Mattila_2010} estimated the mass of the ER, ionized by the initial flash, to be approximately 4.5 and $5.8\times10^{-2}$ $M_{\odot}$, respectively. These estimates were based on emission lines from the ER up to 2000 days post-explosion.

We can estimate the mass and the filling factor of the hotspots by considering the emissivity of  H$\alpha$, $j_{\mathrm{H\alpha}}$. We further consider emission due to cooling by recombination and collisional transitions of H I in a static nebula with a large optical depth. The density dependence of the collisional effects is rather small, which enables us to evaluate the emission for a wide range of densities and temperatures \citep{gaseous_nebulae}. Line ratios of emission from the shocked gas in the ER indicate typical electron temperatures, $T_{\mathrm{e}}$, of $\sim 5000-$50,000$\ \mathrm{K}$ \citep{Groeningsson_2008_02}. For electron temperatures in the range $5000 \ \mathrm{K} < T_{\rm e} <$ 20,000$\ \mathrm{K}$ the emission per unit volume can be written as:
\begin{eqnarray}
\frac{4\pi \ j_{\mathrm{Ha}}}{n_{\mathrm{e}} n_{ \mathrm{p}}} \sim 35 \times 10^{-26} \ \left ( \frac{T_{\mathrm{e}}}{10^{4} \ \mathrm{K}}\right )^{-0.94} \mathrm{erg\ cm^{-3}\ s^{-1}},
\end{eqnarray}
where $n_{\mathrm{e}}$ and $n_{ \mathrm{p}}$ are the electron and proton number densities. The flux, $F$, that we receive at a distance $d=49.6 \ \mathrm{kpc}$ can be written as $F = j_{\mathrm{Ha}}\ V f/ d^{2}$, where $V$ is the emitting volume. Solving for the filling factor gives:
\begin{eqnarray}
\nonumber
f \sim 2.3\times 10 ^{-3} && \left ( 1+2y \right ) \left (  \frac{2\times 10^{16} \ \mathrm{cm^{3}}}{R} \right)^{3} \\
\nonumber
 \times && \left ( \frac{10^{6}\ \mathrm{cm^{-3}}}{n_{\mathrm{e}}}  \right )^{2} \left( \frac{T_{\mathrm{e}}}{2\times10^{4}\ \mathrm{K}} \right)^{0.94} \\
 \times && \left( \frac{F}{5.2\times 10^{-14} \ \mathrm{erg \ cm^{-2}\ s^{-1}}}   \right)
\end{eqnarray}
where $y$ is the ratio of the He number density over H, and $R$ is the radius of the clump. Using hotspot 2 as an example, with a temperature of $T_{\rm e} =$ 20,000$\ \mathrm{K}$ \citep{Groeningsson_2008_02}, a ratio of $y=0.17$ \citep{Mattila_2010}, typical densities of $n_{\mathrm{e}}=10^{6}\ \mathrm{cm^{-3}}$, a maximum flux of $5.2\times 10^{-14} \ \mathrm{erg \ cm^{-2}\ s^{-1}}$ and a radius equal to half its FWHM (see Table \ref{tab:spot size}) results in a filling factor of the shocked gas of approximately $0.3\%$. Note, however, the extreme sensitivity of $f$ to the uncertain density.

The preshock clumps were probably ionized and may give an estimate of the filling factor as well. For the unshocked gas, the highest inferred densities are $3\times10^{4}\ \mathrm{cm^{-3}}$, while the flux of a clump, $F_{\mathrm{preshock}}$, can be estimated to $\sim 3\times 10^{-15} \ \mathrm{erg \ cm^{-2}\ s^{-1}}$ from narrow emission lines of the ER at $\sim 5000$ days \citep{Groeningsson_2008}. The FWHM of spot 2 is evaluated to be $\sim6.0\pm1.0\times 10^{16}\ \mathrm{cm}$, in the first couple of epochs. The filling factor then becomes $f \approx 0.06 \left ( 3\times 10^{16}\ \mathrm{cm}/ R \right )^{3}  \left ( 3\times10^{4}\ \mathrm{cm^{-3}} / n_{\mathrm{e}}\right )^2 $, where the actual radius of the unshocked obstacle is highly uncertain. This shows that the filling factor of the clumps could be small even before being shocked.

The total mass of the optically emitting material can be calculated by:
\begin{eqnarray}
\nonumber
M = 1.4 \times&&10^{-3} \left ( 1+4y \right ) 
\\
\nonumber
\times && \left ( \frac{10^{6}\ \mathrm{cm^{-3}}}{n_{\mathrm{e}}}  \right )
\left( \frac{T_{\mathrm{e}}}{2\times10^{4}\ \mathrm{K}} \right)^{0.94}
\\
\times && \left(  \frac{\sum F_{\mathrm{i}}}{10.5\times 10^{-13} \ \mathrm{erg \ cm^{-2}\ s^{-1}}} \right) \mathrm{M_{\odot}}, 
\end{eqnarray}
where $\sum F_{\mathrm{i}}$ is the sum of the maximum estimated fluxes of the hotspots ($\sim 10.5\times10^{-13} \ \mathrm{erg \ cm^{-2}\ s^{-1}}$). If we use the same typical postshock values for the rest of the parameters, we get a total mass of $\sim 0.24 \times 10^{-2}\ \mathrm{M_{\odot}}$. An alternative method to estimate the mass of the emitting material is to use $M \approx  \rho_{\mathrm{clump}}f \sum V$, where $\sum V$ is the total volume of the hotspots ($\sim 2.6\times 10^{51} \ \mathrm{cm^{3}}$), based on their circularized FWHM. With a filling factor of $0.3\%$, we get a mass of $\sim0.8 \left (  n_{\mathrm{e}} / 10^{6}\ \mathrm{cm^{-3}} \right )\times10^{-2}\ \mathrm{M_{\odot}} $, consistent with our expectations. We emphasize that these are mass estimates of the denser regions within each hotspot and not the total mass of the hotspots or the ER. The estimated mass found in \cite{Mattila_2010} was based on 3 different density components ($1\times~10^{3}$, $3\times 10^{3}$ and $\sim 3\times 10^{4} \ \mathrm{cm^{-3}}$). The measured fluxes most likely correspond to the higher density structure of the ER, which \cite{Mattila_2010} estimated to have a mass of $\sim 1.2 \times 10^{-2}\ \mathrm{M_{\odot}} $ instead. There are also additional spots not included in our calculation whose flux contribution could be considerable. Such spots are for instance those appearing in the western part of the ER at later epochs (after the year 2015) or those in the star region. 

We anticipate an additional mass contribution from the less dense component of the hotspots, where the shocks remain nonradiative and the atoms are collisionally ionized. Before the impact, this component should correspond to gas with the lowest preshock densities in the ER, in the range of 1 to $3\times 10^{3}\ \mathrm{cm^{-3}}$ \citep{Mattila_2010}. Disregarding magnetic field effects, the gas immediately behind the shock front undergoes compression by a factor of $\sim4$ \citep{pun_2002}, and therefore, we can expect postshock densities of $3-12\times 10^{3}\ \mathrm{cm^{-3}}$ for the less dense component. If we further assume a filling factor of $\sim 99.7\%$ for this component, along with the total measured volume of the hotspots, we roughly estimate the contribution of the less dense part to be approximately $\sim 1.2-3.6\times 10^{-2}\ \mathrm{M_{\odot}}$.

Leading theories for explaining the origin and the formation of the ER in SN 1987A include interacting stellar winds from different stages of the progenitor and binary merger \citep{Blondin_1993, Chevalier_1995, Morris_2007, Morris_2009}. While binary merger models have shown success in addressing several constraints associated with the ER, the specific mechanism responsible for the clumping around the ER remains unclear. Some previously proposed formation scenarios include hydrodynamic forces and Rayleigh-Taylor instabilities \citep{Sugerman_2005, Orlando_2015, Zhou_2017_1}. In a recent study, \cite{Wadas_2024} proposed clump formation resulting from Crow instabilities, successfully predicting approximately 30 clumps. Our results give new information about the structure and mass of the spots, which will help assess different models for their formation.

\section{Summary and conclusions}\label{sec: Summary}

The aim of this study was to investigate the characteristics of the hotspots within the ER of SN 1987A, which became visible following the impact of the expanding ejecta with the ER. We tracked their evolution through HST optical imaging observations. A total of 33 images were analyzed, spanning from day 2770 to day 12,978 after the explosion (from year 1994 to 2022). These images were captured using three different instruments (WFPC2, ACS, and WFC3) with the F675W and F625W filters, where the flux is dominated by H$\alpha$ emission from the hotspots. The main findings and conclusions from the analysis are summarized below.

\begin{itemize}

    \item 
    We identified the existence of at least 26 hotspots across the ER, exhibiting a systematic radial expansion from the center of SN 1987A. While additional hotspots became detectable, particularly in the western section of the ER after 10,000 days, their lack of systematic movement led us to exclude them from the analysis. The movement of the hotspots was studied following their complete emergence when their emission could be separated from the diffuse component. The hotspots emerged between 4000 and 7000 days, with those in the northeast appearing first and those in the west appearing last.
    
    \item 
    We studied the geometry of the ER by fitting two different models involving: 1) fitting ellipses to the hotspots, and 2)  fitting the entire ER with an elliptical annulus. Both methods yielded similar results, with an inclination angle of $43.51 \pm 0.38^{\circ}$ ($42.85 \pm 0.50^{\circ}$) and orientation angle of $-7.84 \pm 0.38^{\circ}$ ($-6.24 \pm 0.31^{\circ}$), for model 1 (model 2). All hotspot positions were corrected for projection effects using the 2nd model, allowing for measurements of their full space velocities.
    
    \item 
    We observed a wide range of velocities, with approximately two-thirds of the hotspots displaying at least one break in their expansion rate.  Generally, hotspots exhibited deceleration following the initial interaction with the blast wave. The 90th percentile range of the initial velocities spanned from $390$ to $1660 \ \rm km \ s^{-1}$, with a median of $960\ \rm km \ s^{-1}$. In contrast, the 90th percentile range of subsequent velocities ranged from $40$ to $660 \ \rm km \ s^{-1}$, with a median value of $260\ \rm km \ s^{-1}$. The velocity changes take place between 6700 and 10,600 days, with southeastern spots exhibiting the earliest changes. Only nine hotspots maintained a constant velocity, ranging mainly between $230$ and $830\ \rm km \ s^{-1}$, with a median of $380 \ \rm km \ s^{-1}$.
    
    \item
    Although transmitted shocks into the obstacles can attain velocities of order a few hundred $\rm km \ s^{-1}$, the higher velocities ($\gtrsim 1000\ \rm km \ s^{-1}$) are unlikely to correspond to radiative shock velocities since such velocities would require low preshock densities for the obstacles ($\lesssim 10^4 \ \rm cm^{-3}$) and an extremely long cooling time of order 100 years. The most probable scenario is that we are initially observing a phase velocity resulting from density inhomogeneities within the clumps, coupled with relatively short cooling times of the dense regions (compared to the time for the shock to cross the ER). Specifically, we perceive the hotspot as a collection of small, dense substructures distributed within a less dense diffuse region. As a result, we interpret the initial velocities as a combination of the movement of the shocked substructures and the rate at which these substructures are being shocked.

    \item 
    The velocities observed at later epochs in spots that decelerated should correspond to the actual velocities of the transmitted shocks within the dense substructures. This principle should also apply to spots that did not exhibit any breaks in their expansion rate, as they showed very similar characteristics. These velocities are comparable to typical spectroscopic velocities from H lines in the ER and are consistent with expected velocities of shocks that had time to become radiative. The break times indicate the termination of the initial phase velocity, which most likely occurs when the entire hotspot is shocked. This result is in agreement with previous studies that suggest that the blast wave left the ER $\sim9000$ days after the explosion. 

    \item 
    The light curves of the hotspots exhibited an initial sharp increase in emission, followed by a gradual decline. The peak of the fluxes ranged between $0.95$ and $9.29\times 10^{-14}\ \mathrm{erg\ cm^{-2}\ s^{-1}}$, with spots in the west being the brightest. We found a moderate correlation between the break times of the velocities and the time of maximum brightness, the latter occurring mainly between 6500 and 8900 days, further supporting the existence of substructures within the obstacles.

    \item 
    We found that 13 of the hotspots are spatially resolved. We estimated typical physical sizes between $4$ and $6\times10^{16}\ \mathrm{cm}$, accounting for instrumental broadening effects. We observed no significant systematic increase in size for the resolved hotspots over time in any direction. Resolved spots had an elongation of approximately $0.7-0.8$. Most spots appeared elongated in a direction almost perpendicular to their position angle. This further supports the scenario of a short cooling time, as it could result in emission from a partially shocked clump, making it appear flattened.

    \item
    We estimated a fraction of hotspot volume filled by optically emitting high-density shocked material of $\sim 0.3 \left (  n_{\mathrm{e}} / 10^{6}\ \mathrm{cm^{-3}} \right )^{-2} \%$. For the unshocked clumps, we estimate a filling factor of $\sim 0.06 \left ( 3\times 10^{16}\ \mathrm{cm}/ R \right )^{3}  \left ( 3\times10^{4}\ \mathrm{cm^{-3}} / n_{\mathrm{e}}\right )^2 $. In addition, based on the light curves, we calculate a mass of $ \sim0.24\left (  n_{\mathrm{e}} / 10^{6}\ \mathrm{cm^{-3}} \right )^{-1} \times 10^{-2}\ \mathrm{M_{\odot}}$ for the emitting substructures of the hotspots. A mass estimate using the filling factor and the estimated sizes of the hotspots gives a mass of $0.8\times \left (  n_{\mathrm{e}} / 10^{6}\ \mathrm{cm^{-3}} \right )\times 10^{-2}\  \mathrm{M_{\odot}}$ instead. 
    
\end{itemize}


The HST data presented in this article were obtained from the Mikulski Archive for Space Telescopes (MAST) at the Space Telescope Science Institute. The specific observations analyzed can be accessed via \dataset[DOI: 10.17909/vpxn-d660]{http://dx.doi.org/10.17909/vpxn-d660}.

\begin{acknowledgments}
This research was supported by the Knut \& Alice Wallenberg Foundation. 
\end{acknowledgments}

%

\vspace{5mm}
\facilities{HST (WFC3, ACS, WFPC2)}


\software{Astropy \citep{2013A&A...558A..33A,2018AJ....156..123A}, matplotlib \citep{matplotplib_2007}, Photutils \citep{photutils_1.9}, SciPy \citep{SciPy_2020}, NumPy \citep{numpy}, seaborn \citep{seaborn}
          }


\appendix

\section{Fitting Functions}\label{Appendix: Fitting}

In this appendix, we provide all the functions and models that were used throughout this project. The model used to fit the individual hotspots (Section \ref{subsec:Identification and Fitting of Hotspots}) is identical to the 2D Gaussian function provided by \texttt{Astropy}, with the only exception being that it uses ``Flux" as a parameter instead of ``Amplitude", calculated as:
\[\text{Flux} = 2 \pi \ \text{Amplitude} \ \sigma_x \sigma_y\], 
where $\sigma_x$ and $\sigma_y$ are the spreads of the Gaussian. The expression of the function is given by:
\begin{equation}
\label{eq:2dg}
    f(x, y;\text{Flux},x_0,y_0,\sigma_{x},\sigma_{y},\theta) = \frac{\text{Flux}}{2\rm{\pi} \sigma_{x}\sigma_{y}} \ \exp \left(A \left(x-x_{0}\right)^{2} + B \left(x-x_{0}\right) \left(y-y_{0}\right) + C \left(y-y_{0}\right)^{2}\right)
\end{equation}
where
\begin{equation}
A = -\left(\frac{\cos^{2}\theta}{2\sigma^{2}_{x}} + \frac{\sin^{2}\theta}{2\sigma^{2}_{y}}\right), 
\end{equation}
\begin{equation}
B = 2\left(-\frac{\sin2\theta}{4\sigma^{2}_{x}} + \frac{\sin2\theta}{4\sigma^{2}_{y}}\right), 
\end{equation}
\begin{equation}
C = -\left(\frac{\sin^{2}\theta}{2\sigma^{2}_{x}} + \frac{\cos^{2}\theta}{2\sigma^{2}_{y}}\right).
\end{equation}
Here, $x_0$, $y_0$ is the center of the Gaussian, and $\theta$ is the rotation angle of the blob, measured counter-clockwise. Thanks to its ``Flux" parameter, this function can be used directly as a PSF model with BasicPSFPhotometry or its sub-classes.

The light curves of the spots were fitted using an Exponentially modified Gaussian model whose density function is derived via the convolution of the normal and exponential probability density functions:
\begin{equation}
\label{eq:flux}
    f(t;A,\mu,\sigma,\gamma) = \frac{A \gamma}{2} \ \exp\left( \gamma \left( \mu - t + \frac{\gamma \sigma^{2}}{2} \right) \right) \mathrm{erfc}\left( \frac{\mu + \gamma\sigma^{2} - t}{\sqrt{2}\sigma}\right)  
\end{equation}
Here, the coefficients $A$, $\mu$, and $\sigma$ are the amplitude, center and spread of the Gaussian, $\gamma$ is the rate parameter of the exponential, and erfc() is the complementary error function.

The following expression was used as a model for the ER as a whole and it consists of an elliptical annulus with a Gaussian radial profile centered at $r = R$. The amplitude of the ring is described by two terms: a constant $I_0$ and a sinusoidal component $I_1$, which allows the maximum brightness of the ring to vary as a function of the azimuth. The equation of the elliptical annulus is described by:
\begin{equation}
\label{eq:gaus ring}
    I(x, y; I_0, I_1, \psi_0, x_0, y_0, \theta, \sigma_r, R, \epsilon) = (I_0 + I_1 \ \cos(\psi_0 + \psi)) \ \exp\left(-\frac{(r - R)^2}{2\sigma_{r}^{2}}\right), 
\end{equation}
where
\begin{equation}
     r = \sqrt{\left[\left(x - x_0\right) \cos\theta + \left(y - y_0\right) \sin\theta\right]^2 + \frac{\left[\left(y - y_0\right) \cos\theta + \left(x - x_0\right) \sin\theta\right]^2}{\left(1-\epsilon\right)^{2}}}, 
\end{equation}
\begin{equation}
     \psi = \theta -  \mathrm{atan2}\left(y - y_0, \ x - x_0\right).
\end{equation}
Here, $\psi_0$ is the initial phase of the sinusoidal component, $x_0$, $y_0$ is the centroid of the ring, $\theta$ is the position angle of the major axis measured from the image +y axis, $\sigma_r$ is the width of the Gaussian profile and $\epsilon$ is the ellipticity. 

\section{Inclination Correction}\label{Appendix: Inclination}

In this appendix, we describe the inclination correction that was applied to all of the hotspot positions. The analysis is based on orthographic projection (viewing from an infinite distance), meaning that the major axis of the apparent ellipse will be equal to the diameter of the ER and the center of the ellipse will coincide with the center of the ER. This approximation is sufficient since the radius of the ER is of order $0.2\ \mathrm{pc}$ while the distance to the ER is $\sim 50\ \mathrm{kpc}$. In reality, the relative difference between the semimajor axis and the radius of the ER will be of order $10^{-11}$, while the offset between the apparent center and the center of the ER will be of order $10^{-6}\ \mathrm{pc}$. The deprojection is performed by rotating an ellipse by the inclination angle $i$ about its major axis. The corrected coordinates of the spots are given by:
\begin{equation}
    \left[ \begin{array}{c}
    x_{\mathrm{true}} \\
    y_{\mathrm{true}}
    \end{array} \right]
    = \mathbf{M}
    \left[ \begin{array}{c}
    x_{\mathrm{obs}} \\
    y_{\mathrm{obs}}
    \end{array} \right],
\end{equation}
\begin{equation}\label{matrix}
\mathbf{M} = \frac{1}{\cos i}
\left[ \begin{array}{cc}
    \cos i\ \cos^{2}{\alpha} + \sin^{2}\alpha & \ \sin\alpha\ \cos\alpha(1-\cos i) \\
    \sin\alpha\ \cos\alpha(1-\cos i) & \ \cos^{2} \alpha + \cos i\ \sin^{2}\alpha  \\
\end{array} \right]
\end{equation}
where $x_{\rm{obs}}$, $y_{\rm{obs}}$ is the observed estimated centroid, $i$ is the inclination angle and $\alpha$ is the orientatrion angle. 

\section{Fits to the Radial Movement of Individual Hotspots. }\label{Appendix: Radial Fitting of the Spots}

In this appendix, we provide the plots with the time evolution of the radial distances of the hotspots, starting from the time of their emergence (as defined in Section \ref{sec:Analysis and Results}). These distances were corrected for the inclination of the ER. The plots also include the best-fit model that describes them. 

\begin{figure}[h!]
    \plottwo{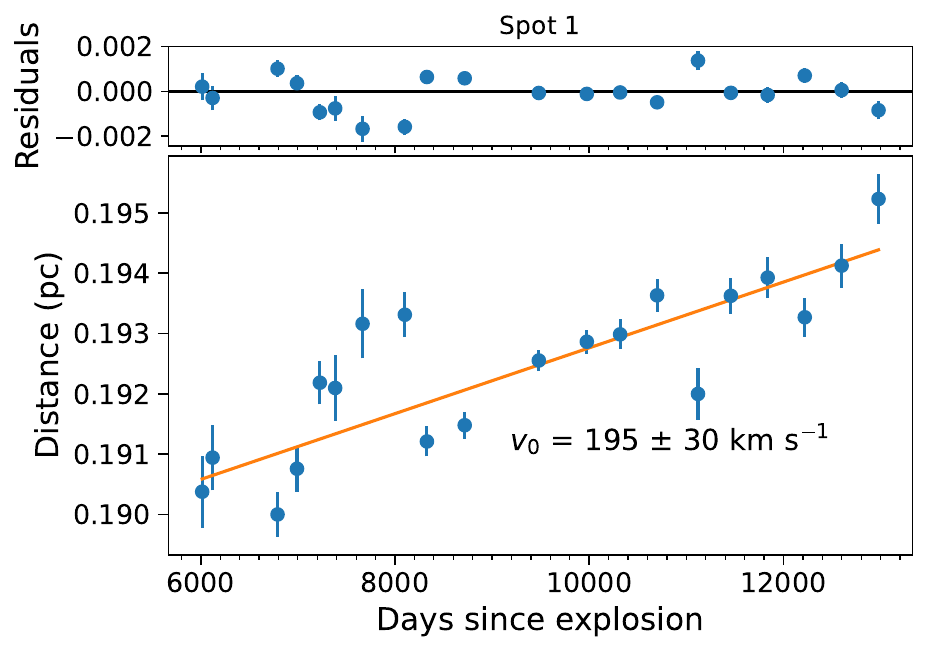}{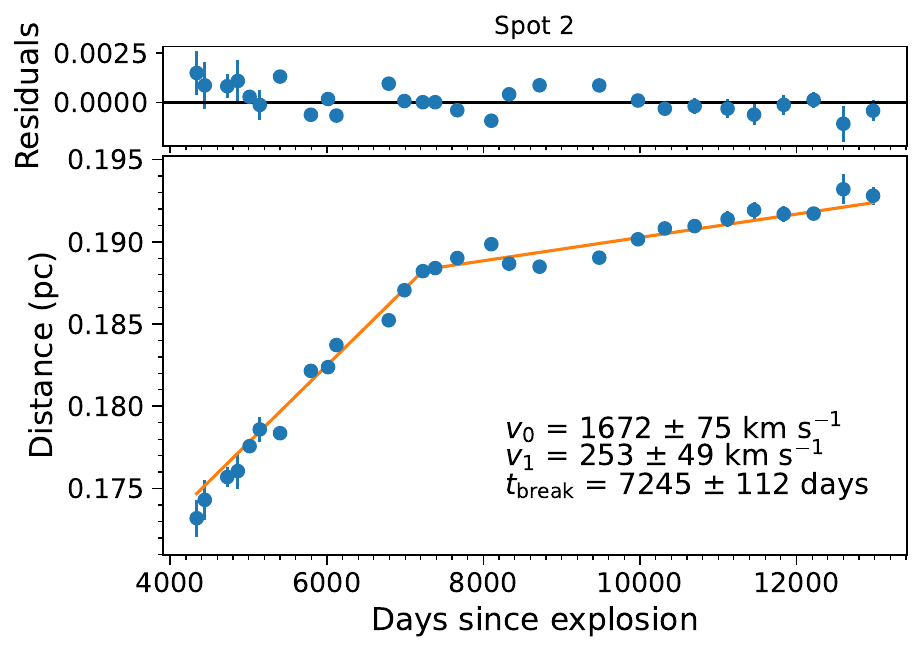}
    \plottwo{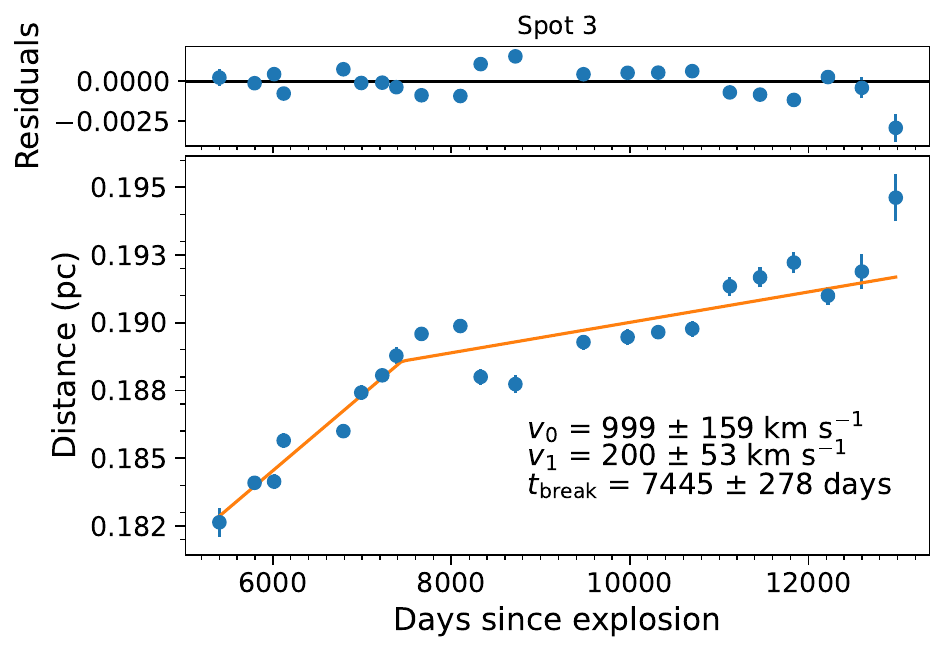}{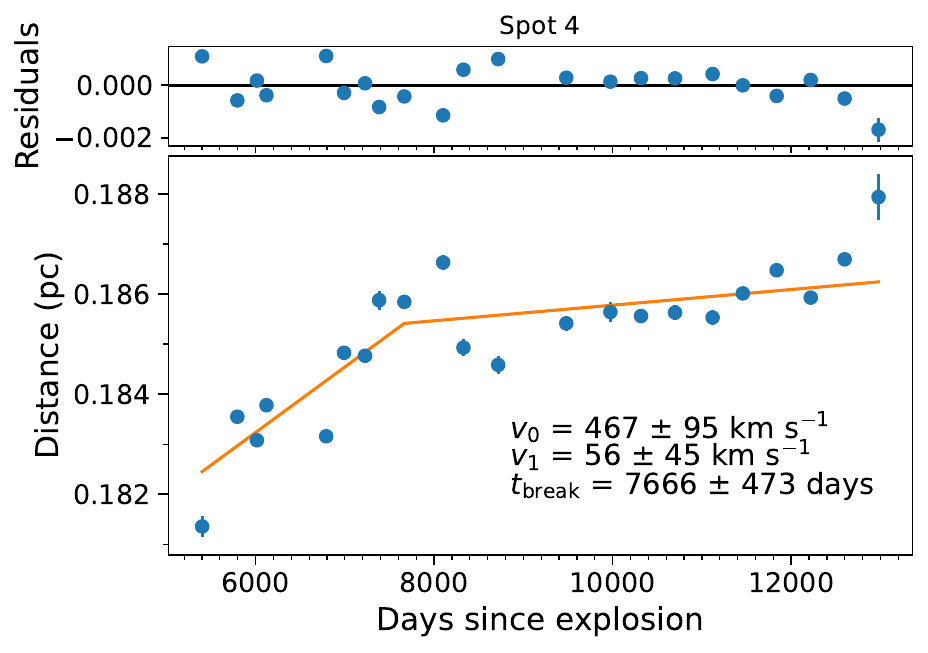}
    \plottwo{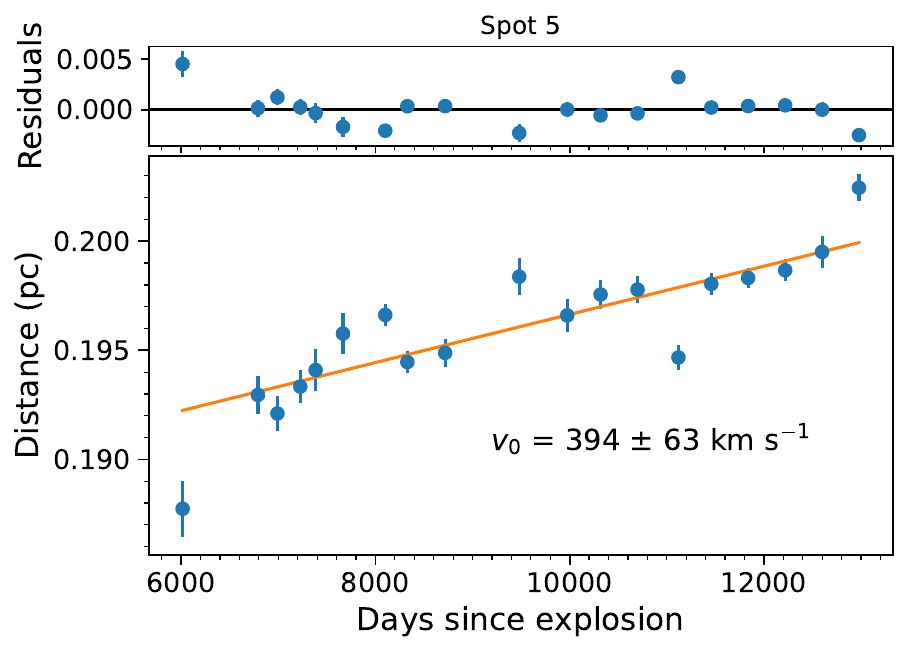}{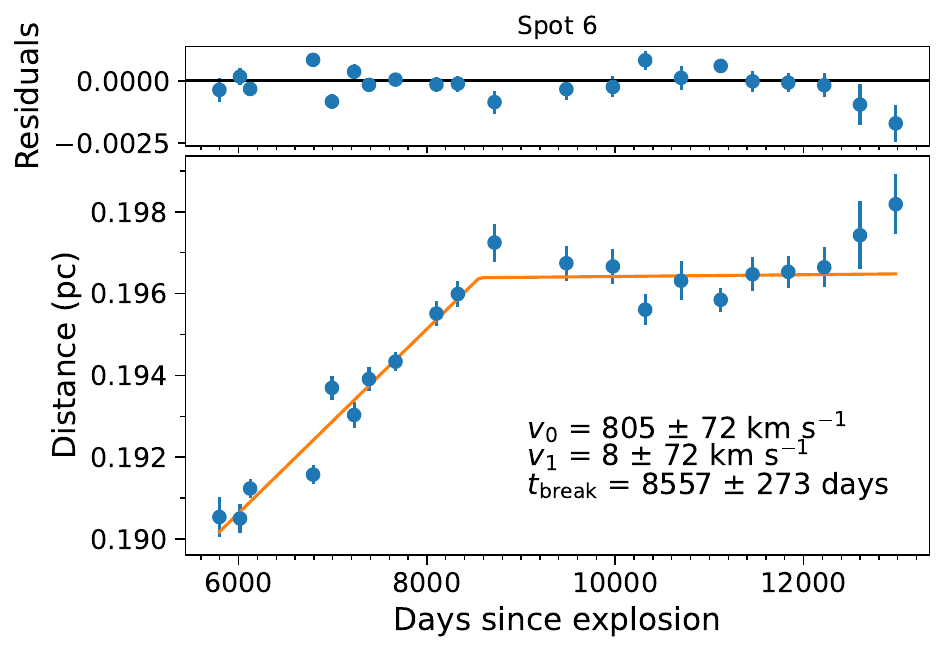}
\end{figure}

\begin{figure}[h]
    \plottwo{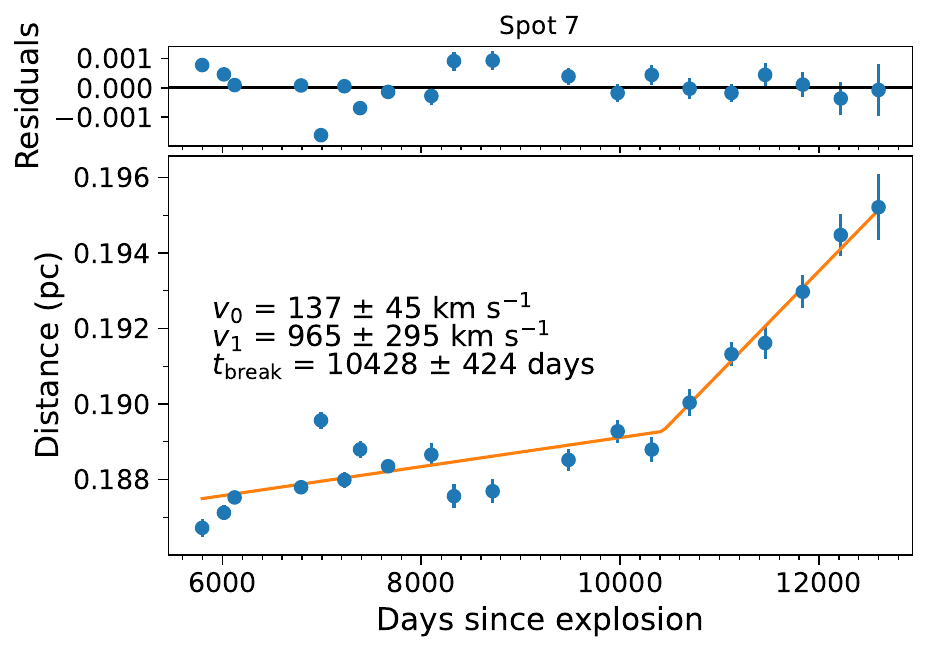}{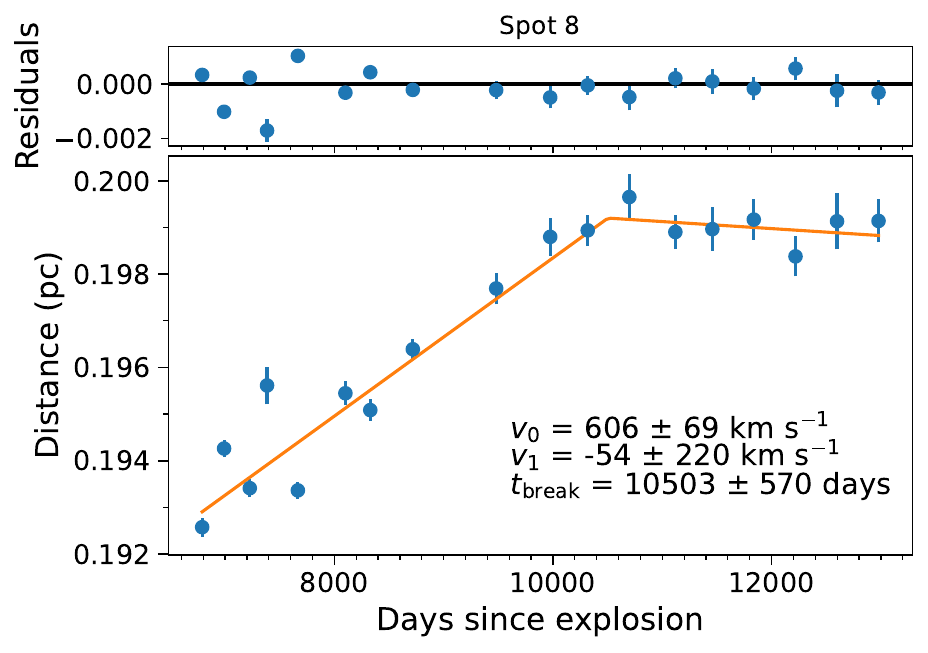}
    \plottwo{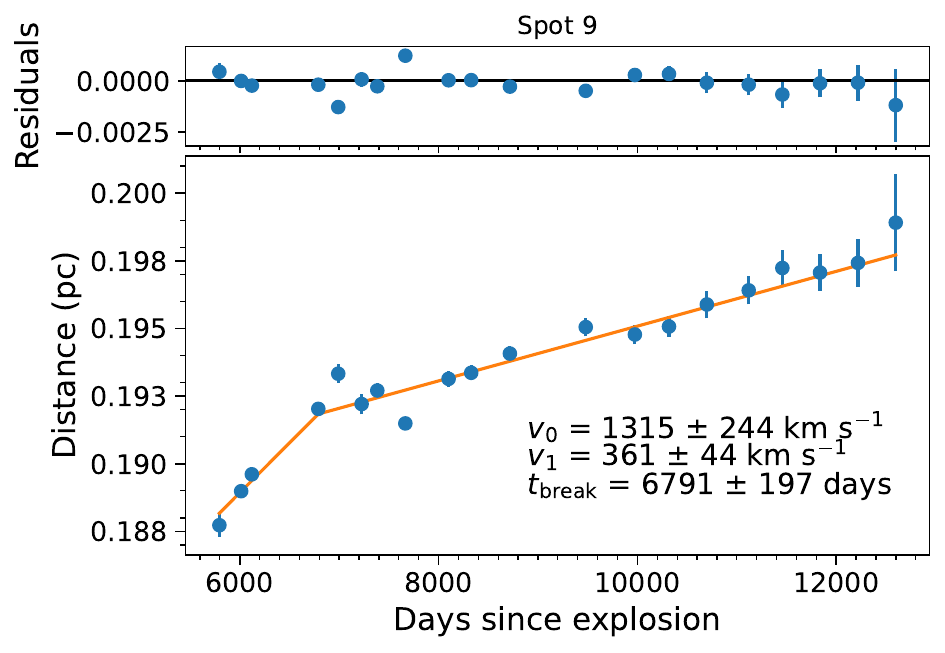}{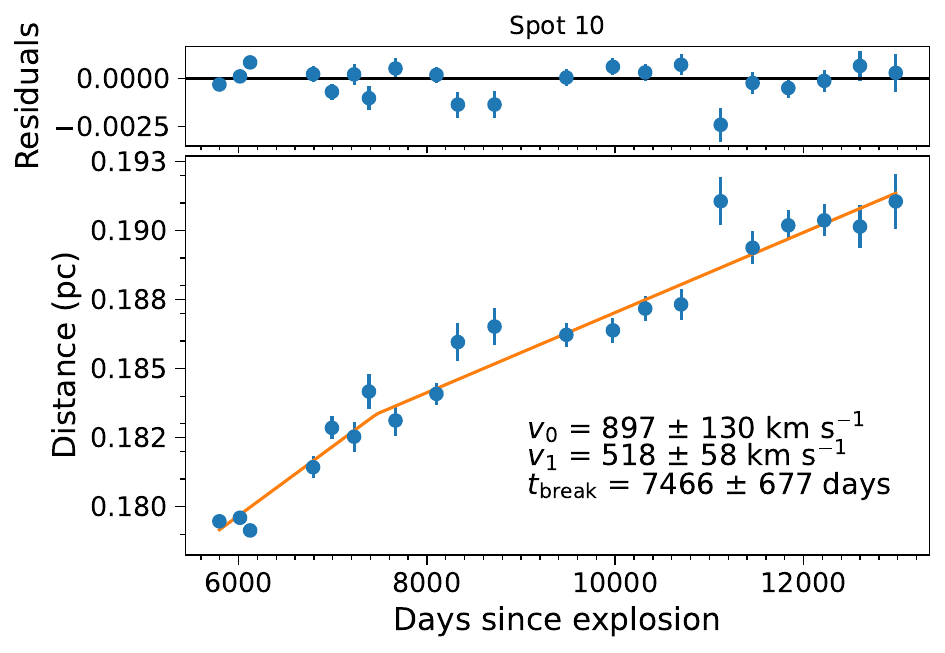}
    \plottwo{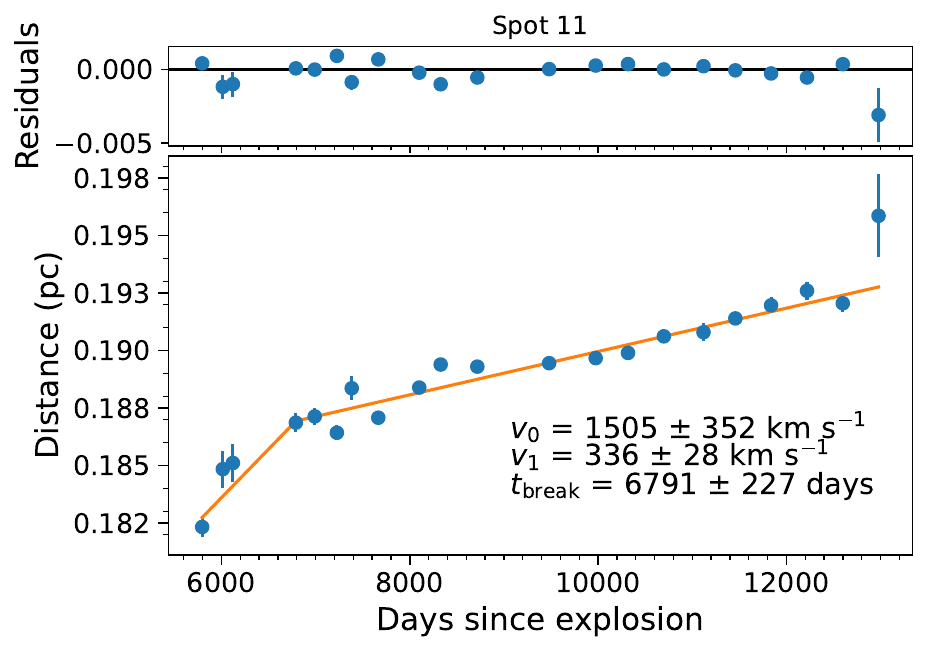}{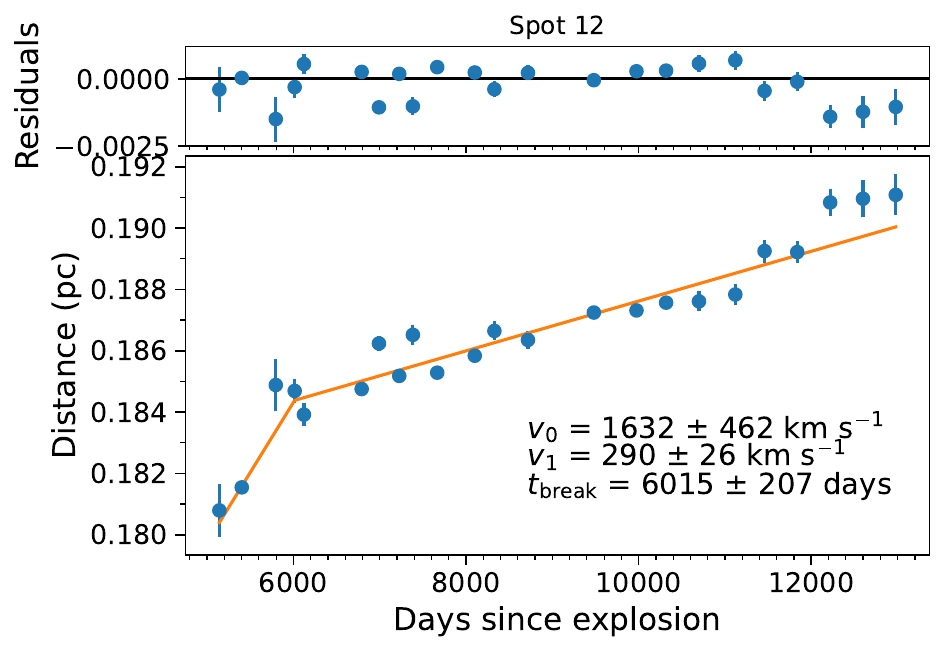}
    \plottwo{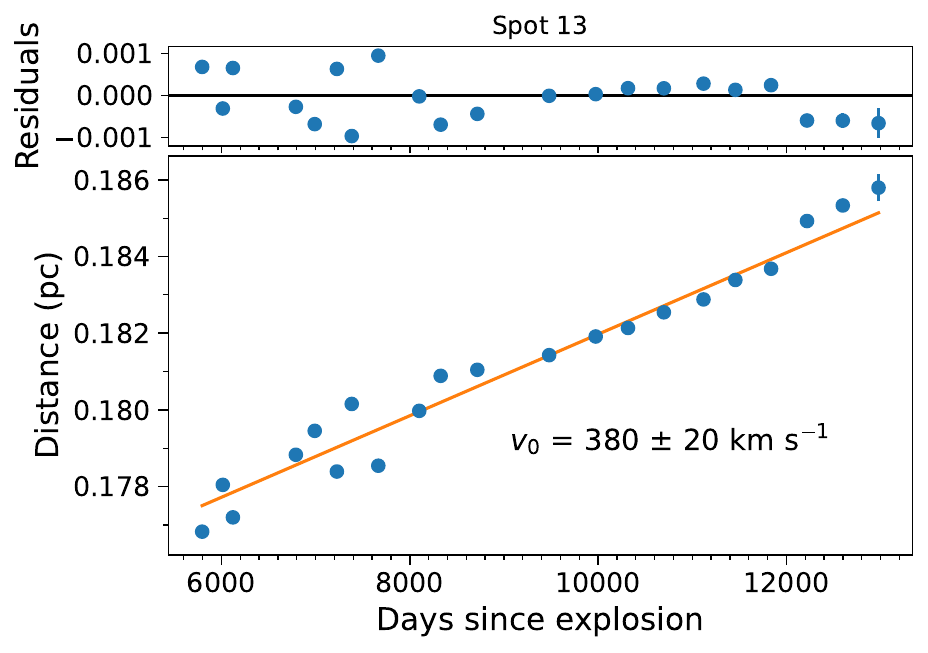}{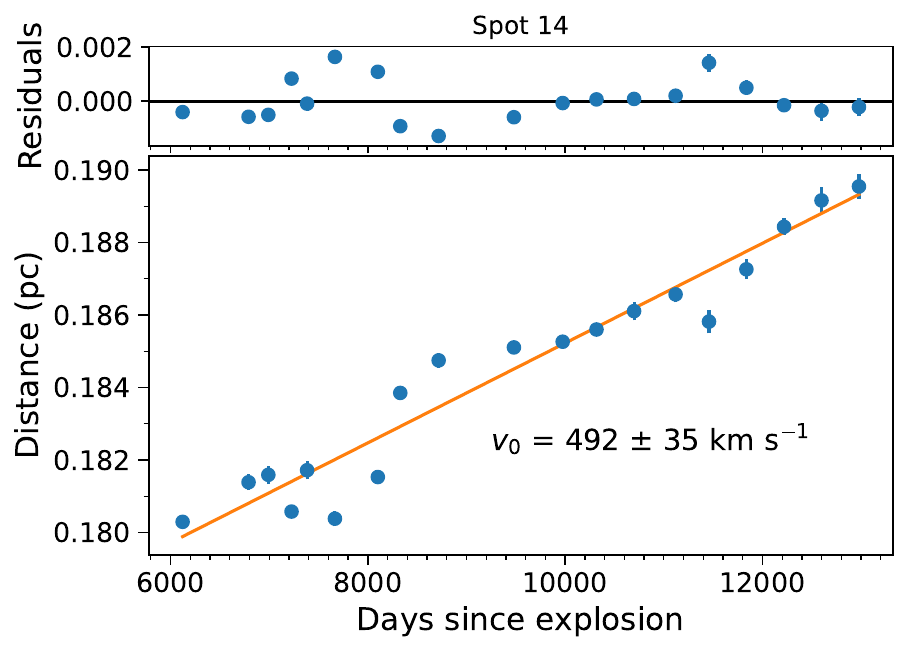}    
\end{figure}

\begin{figure}[h!]
    \plottwo{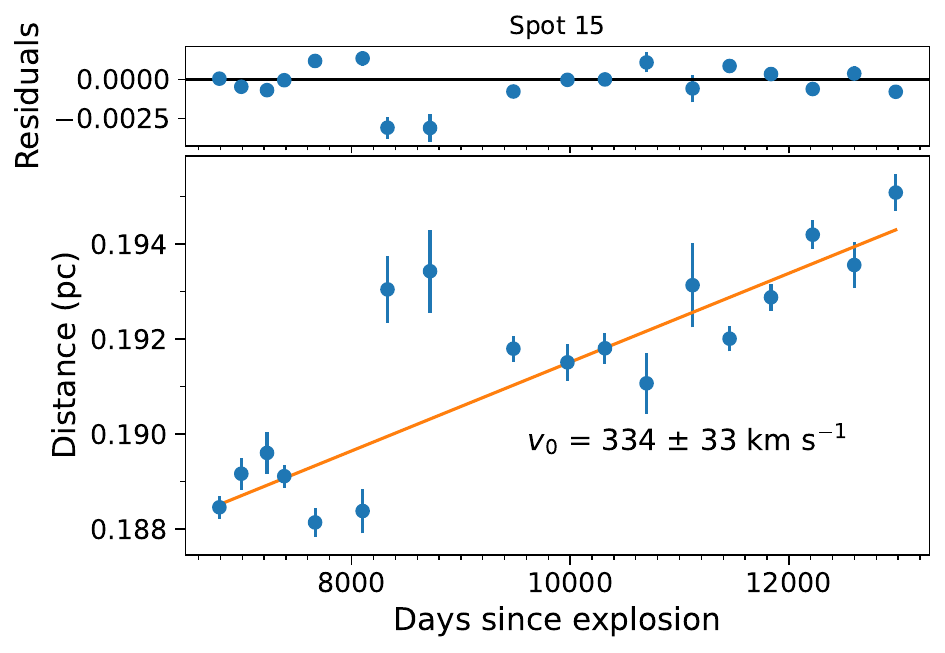}{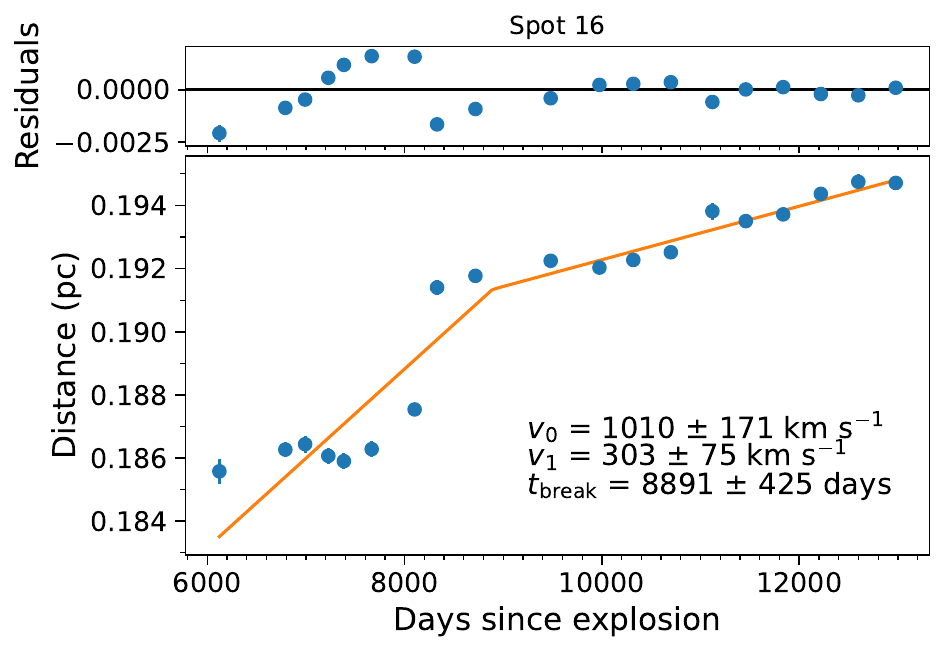}
    \plottwo{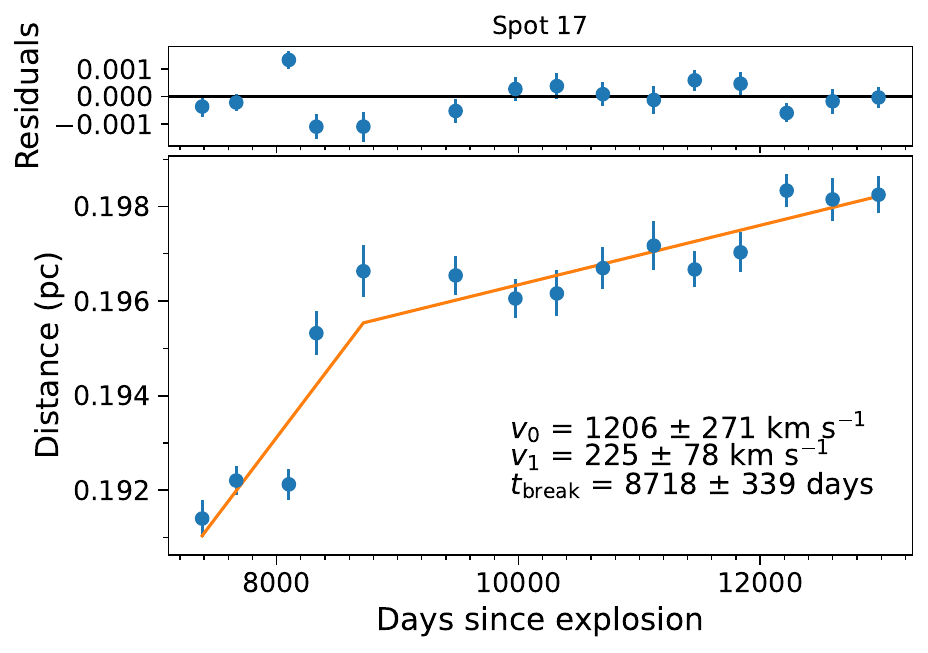}{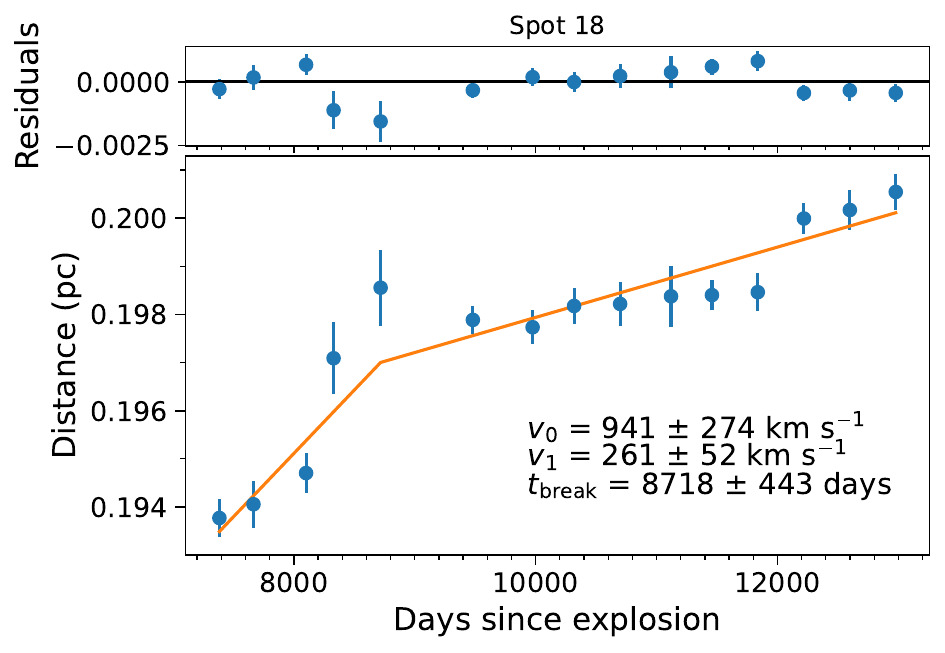}
    \plottwo{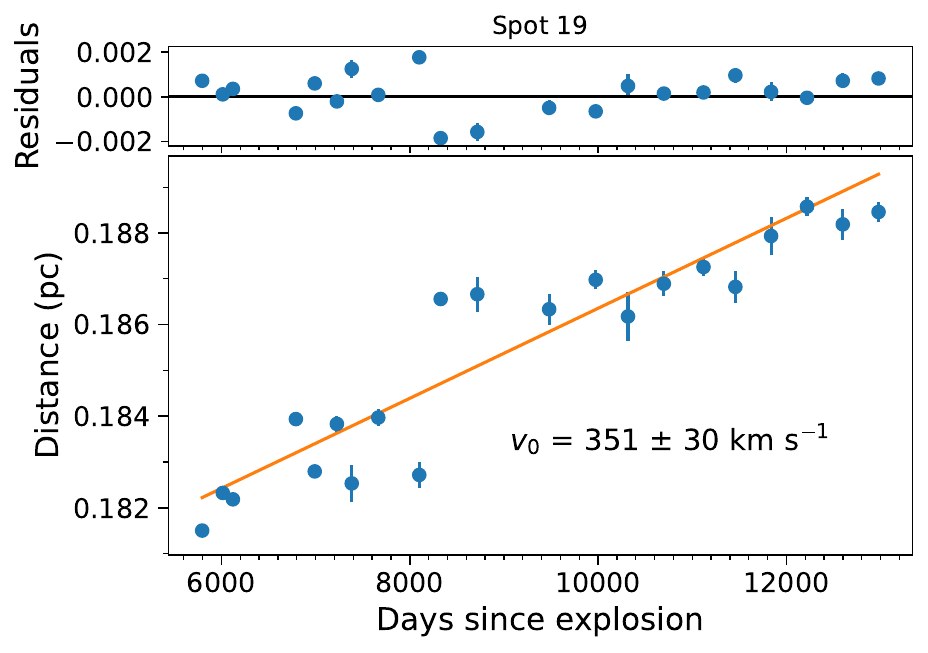}{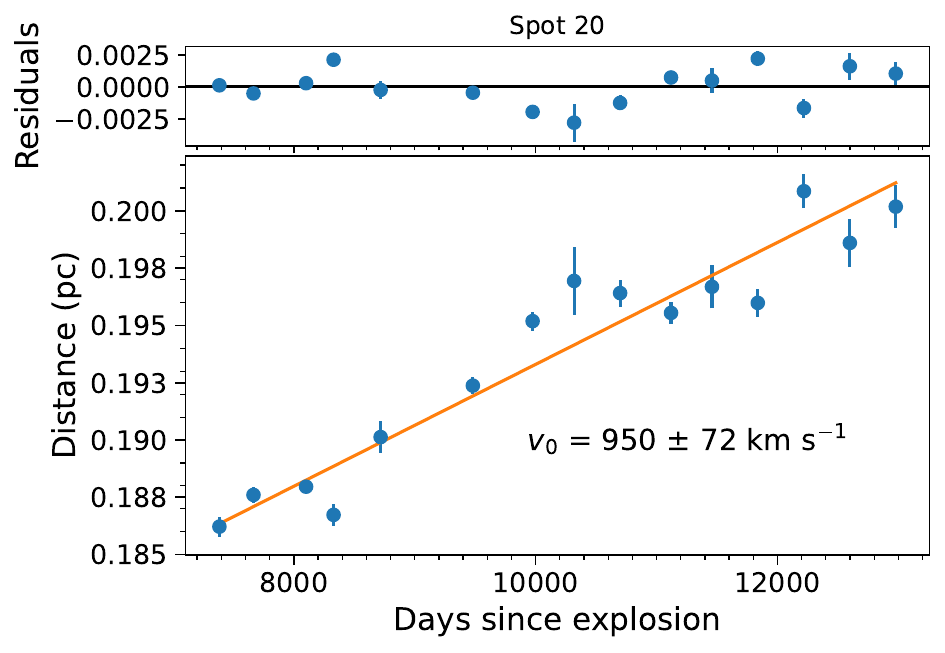}
    \plottwo{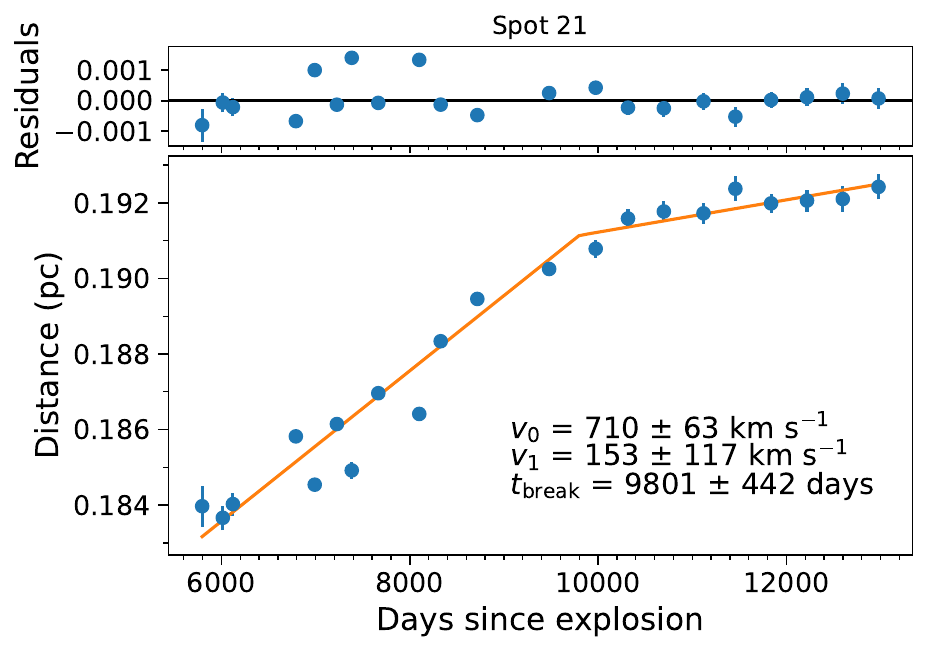}{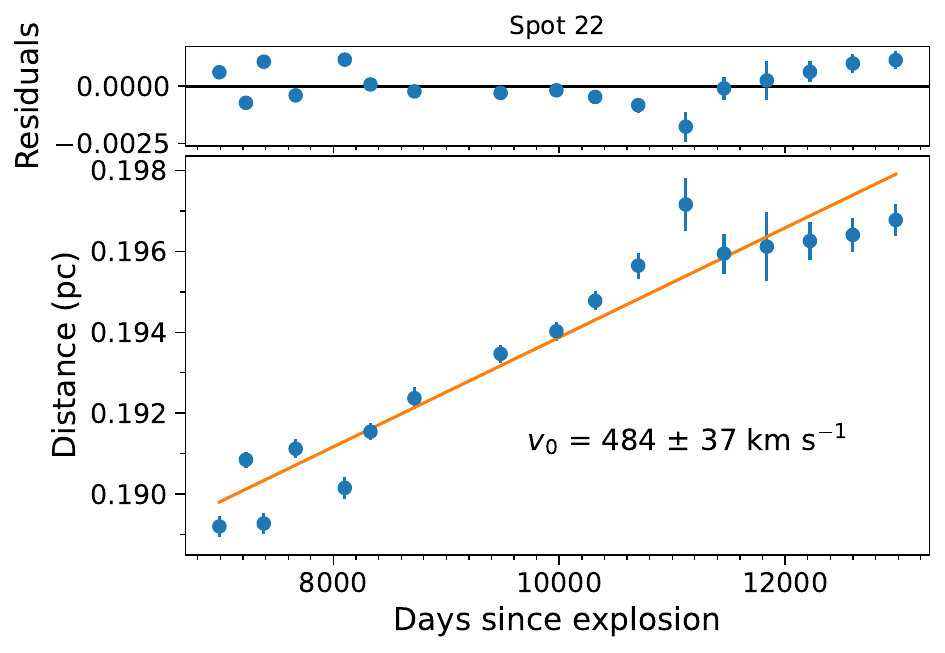}
\end{figure}

\begin{figure}[h!]
    \plottwo{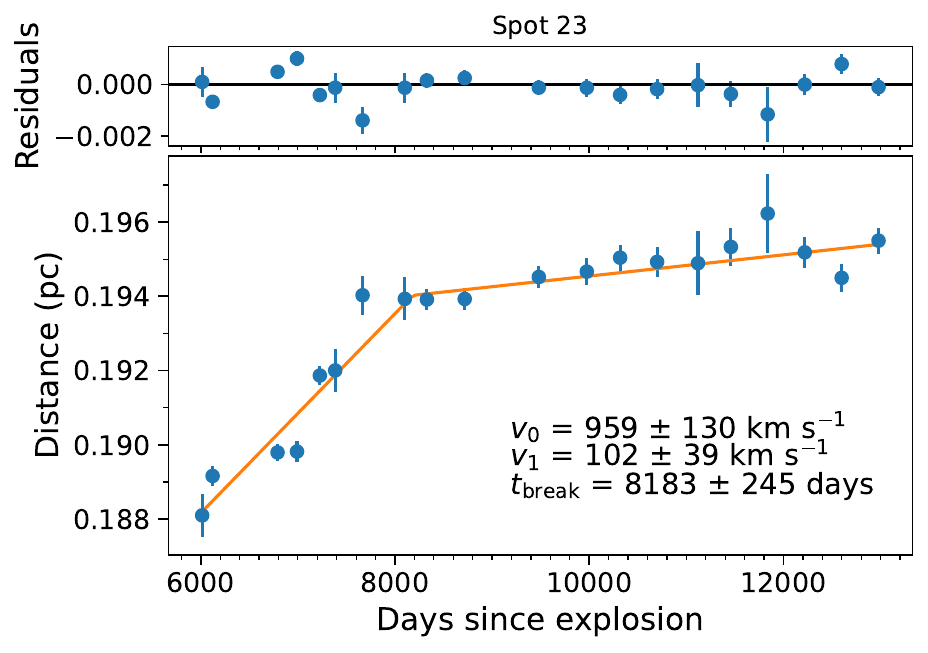}{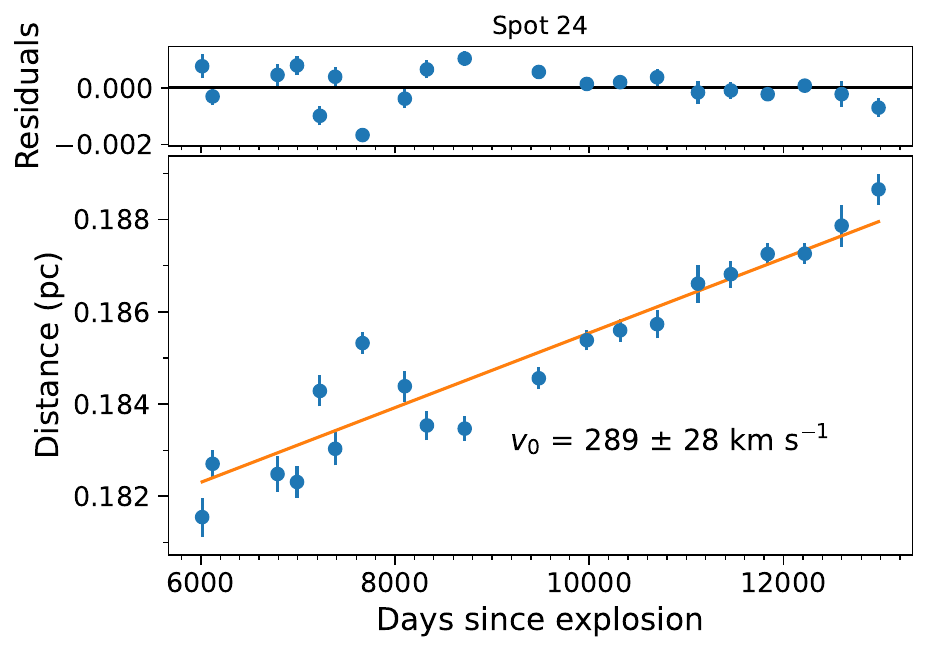}
    \plottwo{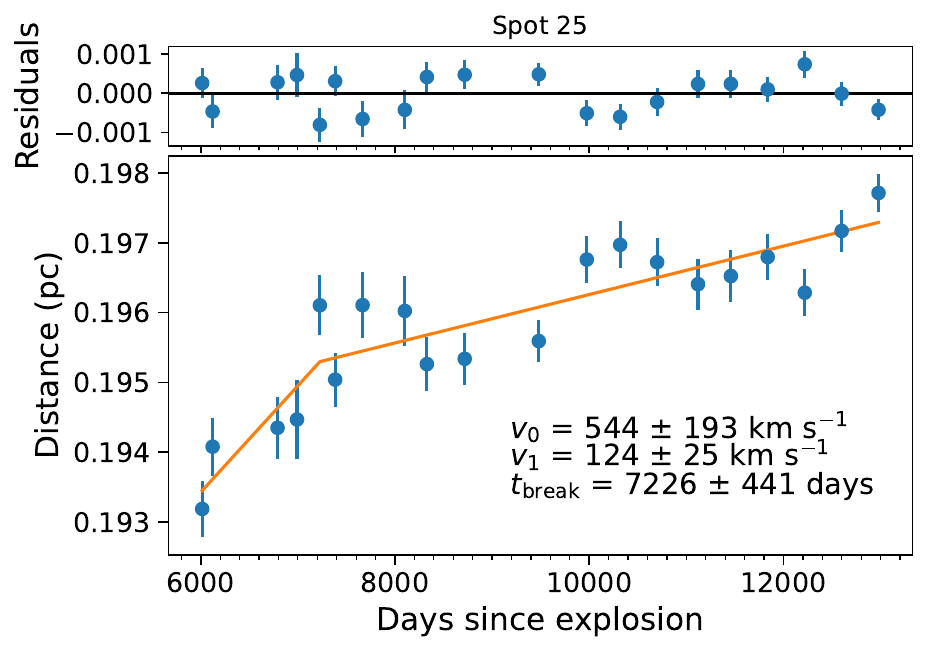}{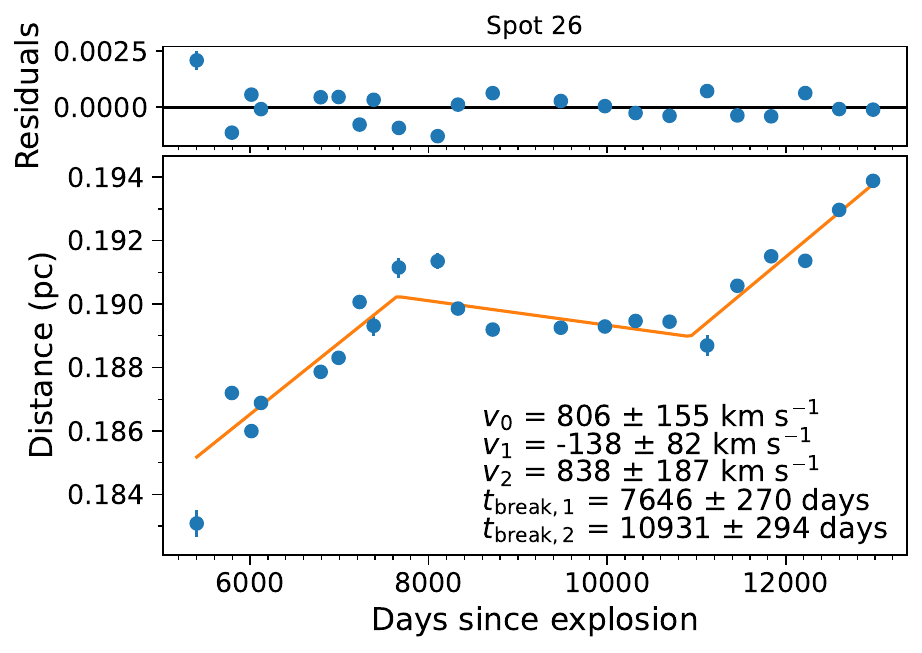}
    \caption{Weighted least square fits of the radial distances of all spots. The residuals shown in the figure are calculated as the weighted difference between the data and the fit, normalized by the uncertainties. \label{fig:best-fits}}
\end{figure}

\clearpage
\bibliography{references}{}
\bibliographystyle{aasjournal}



\end{document}